\documentclass[aps,pra,twocolumn,superscriptaddress,bibnotes,longbibliography]{revtex4-2}%
\usepackage{graphicx}
\usepackage{float}
\usepackage{dcolumn}
\usepackage{bm,color}
\usepackage{amsmath,amssymb,dsfont,amstext,amsfonts}
\usepackage{extarrows}
\usepackage[colorlinks=true,linkcolor=blue,urlcolor=blue,citecolor=blue]{hyperref}
\usepackage{xcolor}
\usepackage{wasysym}
\usepackage{mathtools}
\usepackage{bbold}
\usepackage{lipsum} 
\usepackage{layouts} 

\usepackage[normalem]{ulem} 
\usepackage{color}

\usepackage{multirow} 
\usepackage{tabularx} 

\usepackage[T1]{fontenc} 

\newcommand{\bk}{\bm{k}}

\def\bra#1{\mathinner{\langle{#1}|}}
\def\ket#1{\mathinner{|{#1}\rangle}}

\def\braket#1{\mathinner{\langle{#1}\rangle}}

\def\mod{\:\mathrm{mod}\:}

\begin{document}
\title{Non-periodic Boundary Conditions for Euler Class and Dynamical Signatures of Obstruction}
\author{Osama A. Alsaiari}
\email{oaaa3@cam.ac.uk} 
\affiliation{TCM Group, Cavendish Laboratory, University of Cambridge, JJ Thomson Avenue, Cambridge CB3 0HE, United Kingdom\looseness=-1}
\author{Adrien Bouhon }
\affiliation{TCM Group, Cavendish Laboratory, University of Cambridge, JJ Thomson Avenue, Cambridge CB3 0HE, United Kingdom\looseness=-1}
\affiliation{Nordita, KTH Royal Institute of Technology and Stockholm University, Roslagstullsbacken 23, SE-106 91 Stockholm, Sweden\looseness=-1}
\author{Robert-Jan Slager}
\affiliation{Department of Physics and Astronomy, University of Manchester, Oxford Road, Manchester M13 9PL, United kingdom\looseness=-1}
\affiliation{TCM Group, Cavendish Laboratory, University of Cambridge, JJ Thomson Avenue, Cambridge CB3 0HE, United Kingdom\looseness=-1}
\author{F.~Nur \"Unal}
\email{f.unal@bham.ac.uk}
\affiliation{School of Physics and Astronomy, University of Birmingham, Edgbaston, Birmingham B15 2TT, United Kingdom\looseness=-1}

\affiliation{TCM Group, Cavendish Laboratory, University of Cambridge, JJ Thomson Avenue, Cambridge CB3 0HE, United Kingdom\looseness=-1}


\begin{abstract}
While the landscape of free-fermion phases has drastically been expanded in the last decades, recently novel multi-gap topological phases were proposed where groups of bands can acquire new invariants such as Euler class. As in conventional single-gap topologies, obstruction plays an inherent role that so far has only been incidentally addressed.  We here systematically investigate the nuances of the relation between the non-Bravais lattice configurations and the Brillouin zone boundary conditions (BZBCs) for any number of dimensions. Clarifying the nomenclature, we provide a general \textit{periodictization} recipe to obtain a gauge with an almost Brillouin-zone-periodic Bloch Hamiltonian both generally and upon imposing a reality condition on Hamiltonians for Euler class. Focusing on three-band $\mathcal{C}_2$ symmetric Euler systems in two dimensions as a guiding example, we present a procedure to enumerate the possible lattice configurations, and thus the unique BZBCs possibilities. We establish a comprehensive classification for the identified BZBC patterns according to the parity constraints they impose on the Euler invariant, highlighting how it extends to more bands and higher dimensions. Moreover, by building upon previous work utilizing Hopf maps, we illustrate physical consequences of non-trivial BZBCs in the quench dynamics of non-Bravais lattice Euler systems, reflecting the parity of the Euler invariant. We numerically confirm our results and corresponding observable signatures, and discuss possible experimental implementations. 
Our work presents a general framework to study the role of non-trivial boundary conditions and obstructions on multi-gap topology that can be employed for arbitrary number bands or in higher dimensions.
\end{abstract}
\maketitle

\section{Introduction}
The concepts of topology and obstruction in physics are closely related. It has been known for quite some time that the presence of a Chern number prevents a localized Wannier description~\cite{Thouless84Wannier,Vanderbilt06PRBChern, coh2009}. This guiding idea has recently been the basis for a rather uniform view of band topologies found in topological insulators and metals~\cite{Zhang11RevModPhysTopInsu,Hasan2010_Rev_colloquium,Clas1,clas2,clas3, slager2017review,clas4,clas5, Shiozaki14}. 
Although such single-gap topologies are now accordingly rather well understood, recently novel multi-gap topologies have emerged that do not necessarily need to fit within this paradigm~\cite{Ahn19PRBSymmetry,Bzdušek2019_Science_NonAbelian,Ahn2019_PRX_Failure,Bouhon2020_NatPhys_ZrTe,Adrien20PRB_Geometric,Even_Quench, Palumbo21PRLNonAbelian, Bouhon23Arxivquantum, Bouhon22ArxivMulti, Davoyan24PRB3D, zhao2017PT}. In these systems distinct invariants emerge due to the interplay of multiple bands and band gaps, a prominent example of which is Euler class under a reality condition with the presence of $\mathcal{P}\mathcal{T}$ (parity and inversion) or $\mathcal{C}_2\mathcal{T}$ (two-fold rotation composed with time reversal) symmetry. In such systems band nodes carry non-Abelian charges that, after braiding, can give rise to pairs of bands hosting irremovable nodes that have a finite Euler invariant. While multigap topological physics is a nascent field, it is increasingly being related to metamaterial experiments~\cite{jiang21NatPhysexperimental,Meron_paper,jiang21NatCommfour,Guo21Natureexperimental}, ultracold quantum gases~\cite{Even_Quench, Breach24PRLInterferometry} and various quantum simulators~\cite{EvenQuenchExper,karle2024_arxiv_kicked}, electronic and magnetic settings ~\cite{Bouhon2020_NatPhys_ZrTe,Lee24ArxivEuler,Adrien21PRBTopological,ChenPRB22Non-Abelian}, phonon bands~\cite{Peng22NatCommPhonons,park21NatCommtopological,Gunnar22PRBtopological}, Floquet systems~\cite{Slager24NatCommNon-Abelian} and, lately, interacting systems~\cite{wahl2024exactprojectedentangledpair}.

As might be anticipated, obstruction plays a role also in the context of multi-gap topologies and the Euler class in particular. Since the definition of Euler class necessitates an orientable two-band subspace, this usually dictates an even Euler class~\cite{Bouhon2020_NatPhys_ZrTe}. While meronic phases with odd Euler invariants (as opposed to skyrmionic phases with even invariant) have been reported in certain cases in the presence of Dirac strings or $\pi$ Berry phases~\cite{Meron_paper}, the full extent of the obstruction effects has not been systematically investigated. Concretely, obstruction originates from the non-Bravais lattice configuration and comes at the expense of having a Bloch Hamiltonian that is not periodic at the Brillouin zone boundary~\cite{Simon2020PRBGeometry,Fruchart_2014}. This is represented by a phase factor matrix that is expected to govern the stability and parity of the Euler class invariant, which we show to be closely related to having obstructed states in the system. 

In this paper, we comprehensively investigate the role of 
such phase matrix boundary conditions and especially in the context of meronic Euler phases in two-dimensional three-band systems. In particular, starting from the atomic limit, where atomic sites are fully decoupled with vanishing tunneling, we consider all possible lattice configurations and analyze the interplay of obstruction with multi-gap topological phases. We thus tabulate resulting Euler invariants that can be achieved upon braiding band nodes in different gaps, highlighting also a similar procedure for more bands and higher dimensions. Secondly, we apply these insights within a dynamical context and address the question of observable signatures of this obstruction. While an early manifestation of the Euler invariant was predicted~\cite{Even_Quench} for an even Euler class (in the presence of trivial boundary conditions) 
by employing quantum quenches and paved the way for experimental measurements~\cite{EvenQuenchExper}, such out-of-equilibrium considerations and quench dynamics 
under non-periodic Brillouin zone boundary conditions (BZBCs) and in the odd Euler class case remained completely unexplored.
We show here that the quench signatures in these cases do recover the monopole--anti-monopole pairs but now with them manifesting across multiple Brillouin zones in momentum space, and with the parity of the number of monopoles depending on that of the Euler invariant. 
More specifically, we analytically establish and numerically confirm how the dynamical quench signatures are transformed across adjacent Brillouin zones as imposed solely by the phase matrix BZBCs, presenting their directly observable physical consequences. We also discuss experimental implementations in quantum simulators. 

This manuscript is organized in two parts by first addressing the aforementioned BZBCs and then demonstrating the corresponding dynamical observables following quenches. In Sec.~\ref{Sec:Eulerclass} we start with introducing the Euler class invariant, highlighting the special form it takes for three-band systems. In Sec.~\ref{Sec:OddEu} we review the Bloch formalism for non-Bravais lattice systems and origin of the phase matrix BZBCs generally. Sec.~\ref{Sec:OddEu}B details out a procedure for obtaining the BZBCs when requiring real Hamiltonians for Euler class and Sec.~\ref{Sec:OddEu}C provides our classification for all cases for the three-band Euler systems, proving the allowed invariant parities. In Sec.~\ref{sec:QuenchDynamics} we analytically illustrate the dynamical signatures of Type 2 (obstructed) Euler phases and how they reveal the underlying BZBCs, where we also included a brief recap of quench signatures of the skyrmionic case~\cite{Even_Quench}. We then numerically verify our results for three-band meronic Euler phases in Sec.~~\ref{sec:QuenchDynamics}C, highlighting the effects of non-trivial BZBCs. In Sec.~\ref{Sec:conclusion}, we discuss our results by elaborating on their applicability in the context of non-orientable phases of Euler Hamiltonians, and elucidate on outlines for experimental verification in quantum simulators.

\section{Euler Class}\label{Sec:Eulerclass}

By analogy to the Chern number, two-dimensional real-valued Bloch Hamiltonians may host an Euler invariant $\chi$. The reality of the system requires the presence of $\mathcal{P}\mathcal{T}$ (parity and inversion) or $\mathcal{C}_2\mathcal{T}$ symmetry (two-fold rotation and time reversal), which we assume in the rest of this work. The Euler class then arises within a two-band subspace (a set of two bands) and, hence, involves the non-Abelian Berry curvature of these bands. Concretely, the value of the Euler invariant $\chi_{i,i+1}$, residing between bands $i$ and $i+1$ can be calculated as, 
\begin{equation} \label{Eq:FullEuClass}
    \chi_{i,i+1}(\mathcal{D})\! =\!\frac{1}{2\pi}\! \left[\int_{\mathcal{D}} dk^2\ \text{Eu}_{i,i+1}(\bm{k}) -  \oint_{\partial\mathcal{D} } \bm{A}_{i,i+1}(\bm{k}) \cdot d\bm{k}\right],
\end{equation}  
by integrating the Euler curvature over a patch $\cal{D}$ in the Brillouin zone (BZ) combined with a loop integral of the Euler connection over the contour of the patch $\partial \mathcal{D}$~\cite{Ahn2019_PRX_Failure,Bouhon2020_NatPhys_ZrTe}.
In the above, we have defined the Euler curvature  two-form $\text{Eu}_{ij} (\bm{k}) = \left\langle \partial_{k_x} u_i(\bm{k}) | \partial_{k_y} u_j(\bm{k}) \right\rangle - \left\langle \partial_{k_y} u_i(\bm{k}) | \partial_{k_x} u_j(\bm{k}) \right\rangle $ and the Euler connection one-form  $\bm{A}_{ij}(\bm{k}) = \bra{u_i(\bm{k})}\bm{\nabla} \ket{u_j(\bm{k})} $ for the Bloch states $\ket{u_i(\bm{k})}$. Note importantly that the Euler forms mix two distinct states as opposed to the Chern forms. 

The Euler class, physically speaking, 
characterizes the topological obstruction to annihilate the band nodes harbored in the two-band subspace $\left\{i,i+1 \right\}$~\cite{Ahn2019_PRX_Failure,Bouhon2020_NatPhys_ZrTe}. While normally two generic band nodes between two bands might be thought to trivially cancel each other (i.e.~gap out) upon combining in the same way that they are created from vacuum since the total topological charge must be conserved, Euler systems beat this conventional notion by involving non-Abelian braiding of band singularities between different subspaces. That is, the band nodes carry non-Abelian charges~\cite{Bzdušek2019_Science_NonAbelian,Breach24PRLInterferometry} in Euler materials, and braiding can result in similarly charged topological nodes within a subspace that can no longer gap out, which is exactly what is measured by the non-zero valued Euler class~\eqref{Eq:FullEuClass}~\cite{Bouhon2020_NatPhys_ZrTe}.

While the definition of multi-gap topological Euler class is completely general, i.e.~for any two-band subspace within a multi-band context, three-band systems separated into two-band (e.g.~$\left\{i,i+1 \right\}=\left\{2,3\right\}$) and single-band ($i=1$) subspaces, hereafter referred to as single gap three-band systems, correspond to classifying spaces that comprise the real projective plane $\mathbb{RP}^2$ (namely the space of unsigned directions in three dimensions). As a result, the Euler form of the two-band subspace can be written in terms of the real vector $\bm{n}(\bm{k})= \bm{u}_2(\bm{k}) \times \bm{u}_3(\bm{k}) \in S^2$ (a point of the unit sphere)~\cite{Even_Quench}, that is the cross product of the eigenvectors of the Euler subspace. Given that the eigenvectors of a real Hamiltonian form an orthonormal frame in $\mathbb{R}^3$\cite{Bouhon2020_NatPhys_ZrTe, Even_Quench}, this corresponds to the eigenvector of the gapped band of the system $\bm{n}(\bm{k}) \equiv \bm{u}_1(\bm{k})$. 

Additionally, considering the simple case of three-band Bloch Hamiltonians which are periodic over a single BZ independently of the chosen Fourier gauge (see next section for detailed definition), it can be shown that the loop integral in Eq.~\eqref{Eq:FullEuClass} vanishes over the boundary of the whole BZ, defined by $\partial \text{BZ} = \ell_2^{-1} \circ \ell_1^{-1} \circ \ell_2 \circ \ell_1 $ (where $\ell_{1,2}$ are the one-dimensional CW cells~\cite{Adrien20PRB_Geometric}, i.e.~loops, of the flat torus), since the Euler connection$-$in $SO(3)$$-$is commutative (c.f.~Section \ref{subsubsec:EulerConstraint} where we discuss the non-periodic BZ case). Consequently, in a three-band system, the Euler curvature takes the form 
 $ \bm{n}(k) \cdot (\partial_{k_1}\bm{n}(\boldsymbol{k}) \times \partial_{k_2}\bm{n}(\boldsymbol{k}))\,$~\cite{Bouhon2020_NatPhys_ZrTe}, 
thus leading to the Euler class
\begin{align} \label{Eq:EuClass_n_Winding}
    \chi_{2,3} = & \frac{1}{2\pi} \int_{\text{BZ}} dk^2 \, \bm{n}(\bm{k}) \cdot (\partial_{k_1}\bm{n}(\boldsymbol{k}) \times \partial_{k_2}\bm{n}(\boldsymbol{k})) \\ 
     & \in \,\{0,\pm2,\pm4...\}
     \equiv 2\mathbb{Z}\,. \nonumber
\end{align}
Note the missing factor 1/2 in the Euler curvature as compared to the Chern form, which shows that one adds $2$ to $\chi_{2,3}$ each time $\boldsymbol{n}$ wraps the unit sphere once. Namely, the invariant over a closed manifold physically amounts to twice the winding of $\bm{n}(\bm{k})$ over the unit sphere~\cite{Bouhon2020_NatPhys_ZrTe}. Noting that the wrapping of a unit vector is integer quantized over a closed manifold, $\chi_{2,3}$ is thus an even integer for three-band Bloch systems with periodic BZBCs, which has remained at the focus in the literature so far. In the following, we present a comprehensive characterization of non-periodic BZ boundaries in the context of three-band real Bloch Hamiltonians in two dimensions, including a detailed analysis of the boundary condition effects on the Euler invariant and enumeration of all possible cases with possible invariants.

\section{Parity Classification of Three-band Euler Systems} \label{Sec:OddEu}

We begin by considering three-band (real) Euler Hamiltonians in two dimensions to address the classification of multigap topologies with general non-trivial BZBCs which are associated to Zak's choice of the Fourier gauge. The non-triviality of a BZBC stems from having odd numbers of atomic orbitals located on $\mathcal{C}_{2}$ (or inversion) symmetric sub-lattice sites belonging to the boundary of the unit cell, as will be shown in detail in the subsequent.
In particular, we demonstrate that non-trivial BZBCs correspond to a rotation of eigenvectors by a phase matrix $V$ due to the shifts in the Wyckoff positions~\cite{Ahn2019_PRX_Failure,Meron_paper} and the Euler class is still well defined and integer quantized, but with a parity that depends on the form of $V$. In Sec.~\ref{subSec:Eu_Parity_Classification}, we present a classification of all possible BZBCs that three-band two-dimensional Euler systems can exhibit, along with the corresponding parity of the Euler class for each case, as summarized in Table. \ref{Tab:classification}. We emphasize that our analysis can be readily extended to higher number of bands and in fact to higher dimensions following a similar procedure.

We highlight that in this manuscript, we go beyond the previously considered cases of even (skyrmionic) Euler phases~\cite{Even_Quench,EvenQuenchExper}, and investigate the effects of non-periodic boundary conditions allowing for odd (meronic) Euler phases~\cite{Meron_paper}. The naming of \textit{skyrmionic} and \textit{meronic} Euler phases is motivated by the fact that Eq.~\eqref{Eq:EuClass_n_Winding} still holds in the presence of non-trivial BZBCs but with $\chi_{2,3} \in \mathbb{Z}$ in general. Consequently, we show that the parity of the Euler class indicates if $\bm{n}(\bm{k})$ wraps the unit sphere integer or half integer times per BZ with the latter still being quantized, i.e.~is skyrmionic or meronic respectively. 

\subsection{Review of Zak's Gauge \& Brillouin Zone Boundary Conditions for Real Hamiltonians}

It is useful to start with a review of the relation between the crystal lattice structures in two dimensions and the BZBCs of their Bloch Hamiltonians to fix the nomenclature, and in particular in the case of real Hamiltonians such as those possessing  $\mathcal{C}_2\mathcal{T}$ symmetry. We present the canonical (natural) gauge choices that lead to a physically motivated Zak phase definition following Ref.~\cite{Fruchart_2014} along with detailed derivations in Appendix~\ref{ap_blochframe}. Our approach is constructive, namely, by deriving the properties of the Bloch Hamiltonian by setting the configuration of the atomic degrees of freedom populating a given crystal system. We then address the question of characterizing a Bloch Hamiltonian that would be given without the knowledge of the corresponding atomic/orbital degrees of freedom. 
Moreover, as part of our procedure for identifying the canonical gauge, we emphasize the significance of the \textit{periodictization} of the Bloch vector field, that is choosing gauge choices that give the "as-periodic-as-possible" form of the Bloch basis, since the key quantities that characterize the Bloch Hamiltonian are obtained from parallel transport within the Bloch vector field.


Considering lattice systems with atomic sites at $\left\{\bm{r}_i\right\}_{i = 1,2..., M}$ (this includes sublattice and orbital degrees of freedom, c.f.~Appendix~\ref{ap_blochframe}) such that the lattice possesses  $\mathcal{C}_{2}$ symmetry ($\pi$-rotation around the vertical $\hat{z}$-axis crossing the center of the unit cell that is perpendicular to the plane) and time reversal symmetry $\mathcal{T}$, the combined $\mathcal{C}_2\mathcal{T}$ symmetry constitutes an anti-unitary symmetry that leaves the 2D momentum unchanged (since both $\mathcal{C}_2$ and $\mathcal{T}$ take $\boldsymbol{k}$ to $-\boldsymbol{k}$) and squares to identity, $[\mathcal{C}_2\mathcal{T}]^2=+1$ \footnote{This condition is satisfied both for bosonic $\mathcal{T}^+$ and fermionic $\mathcal{T}^-$ time reversal symmetries, where $[T^{\pm}]^2 = \pm 1$. Such a condition is also satisfied by $\mathcal{P}\mathcal{T}^+$ symmetry with $\mathcal{P}$ inversion symmetry and $\mathcal{T}^+$ the bosonic time reversal symmetry.}. More precisely, we consider lattices where the atomic sites are at a \textit{$\mathcal{C}_2$ center}, i.e.~sites that map back to themselves upon a $\mathcal{C}_2$ transformation, and leave more generic $\mathcal{C}_2$-symmetric lattices to Appendix~\ref{appsec:A_c2t_and_reality} where we demonstrate that the physical concepts discussed in the main text still hold. It follows that a $\mathcal{C}_2\mathcal{T}$ symmetric system can be written in a basis in which the Bloch Hamiltonian $H(\bm{k})$ is real, where such a basis can generally be found using Takagi factorization~\cite{Bouhon2020_NatPhys_ZrTe} as we demonstrate in Appendix~\ref{ap_blochframe}. For concreteness, we assume in the following that the system is symmetric under both $\mathcal{C}_2$ and $\mathcal{T}^+$. 

From the spectral decomposition of the real Bloch Hamiltonian matrix
\begin{equation}
    H(\boldsymbol{k}) = \mathcal{R}(\boldsymbol{k}) \cdot \mathcal{E}(\boldsymbol{k})\cdot \mathcal{R}(\boldsymbol{k})^{\top},
\end{equation}
where $\mathcal{R}(\boldsymbol{k})$ is the orthonormal frame of real eigenvectors $\mathcal{R}(\bm{k}):\left\{ \bm{u}_1(\bm{k}),\bm{u}_2(\bm{k}),\bm{u}_3(\bm{k}),...,\bm{u}_M(\bm{k}) \right\}$ and $\mathcal{E}(\boldsymbol{k})$ is the diagonal matrix of eigenvalues, we derive that:
\begin{align}\label{eq:BZBC&Vmat_defn}
    \mathcal{R}(\boldsymbol{k}+\boldsymbol{G}) &= V^\dagger(\boldsymbol{G})\cdot \mathcal{R}(\boldsymbol{k})\, &\nonumber\\
    \implies\bm{u}_n(\bm{k} + \bm{G}) &= s_n V(\bm{G})\  \bm{u}_n(\bm{k}), \ \ n = 1,2,3...,M \  &\nonumber \\
    \text{s.t.}\ V(\boldsymbol{G}) &= \text{diag}\left(
         e^{\text{i} \boldsymbol{G}\cdot \boldsymbol{r}_{\alpha_1} },
         \cdots
         , e^{\text{i} \boldsymbol{G}\cdot \boldsymbol{r}_{\alpha_M} } 
         \right)  ,
\end{align}
such that $V(\bm{G})$ is a real orthonormal diagonal matrix (up to an overall phase) as a consequence of picking the Zak Fourier gauge~\cite{Fruchart_2014} (see Appendix~\ref{ap_blochframe}  for details) and  $s_n=\pm 1$ is a free gauge sign. This transformation of the Bloch eigenvectors upon reciprocal lattice translations [$\boldsymbol{G}=n_1\bm{b}_1+n_2\bm{b}_2$, $n_1,n_2 \in \mathbb{Z}$] to adjacent Brillouin zones is what we refer to as \textit{Brillouin zone boundary conditions}. Moreover, along each of the reciprocal primitive lattice vectors $\bm{b}_1$ and $\bm{b}_2$ in two-dimensional (2D) $\bm{k}$-space, there can be in general different BZBCs, $V(\bm{b}_1)$ and $V(\bm{b}_2)$ respectively, which are in turn determined by the lattice configuration of the system.

We note that the periodic gauge choice $s_n = 1\ \forall\bm{k}\ \forall n$ is the canonically adopted choice to avoid introducing an additional $\pi$ phase at the BZ boundary and thus allow for a physically motivated definition~\cite{Fruchart_2014} of the Zak phase~\cite{Zakphase},
\begin{equation}\label{Eq:Zak_phase}
    \gamma_n (\bm{b}_j) = i \oint_{\bm{b}_j}^{} d\bm{k} \cdot \bra{\bm{u}_n(\bm{k})}\nabla_{\bm{k}} \ket{\bm{u}_n(\bm{k})} ,\  j=1,2
\end{equation}
where $\bm{b}_j$ are the reciprocal lattice directions and, to correctly compute these phases, the parallel transport gauge is chosen while enforcing the BZBC for $\bm{u}_n(\bm{k}+\bm{b}_j)$.

\subsection{BZBCs and Stability of Euler Topology for Three-band Systems}

We now turn to classifying all possible configurations for three-band Euler systems, and discussing the effects of the corresponding BZBCs on the Euler class invariant.

\subsubsection{Enumeration of Possible Lattices \& Corresponding BZBCs}\label{subsubsec:eumeration-of_possible_lattices}

Given the symmetry imposed reality of the Hamiltonian matrix $H(\bm{k})$, we have deduced that $V(\bm{G})$ is real, but we note that this holds up to an overall complex phase. We show in Appendix~\ref{sec:BLorigin_gauge} that this global phase of $V(\bm{G})$ corresponds to the gauge choice of setting the origin of the Bravais lattice (BL) relative to the atomic orbital sites. In particular, using the relation between the lattice vectors $\bm{a}_i \cdot \bm{b}_j = 2\pi \delta_{ij}$, we find that the BL origin choices which directly give a real $V(\bm{G})$~\eqref{eq:BZBC&Vmat_defn} correspond to orbital positions that lie at one of the four $\mathcal{C}_2$ center sites $\bm{r}_i = n_{1,i}\frac{\bm{a}_1}{2} + n_{2,i}\frac{\bm{a}_2}{2}$ for $n_{1,i},n_{2,i} \in \left\{0,1 \right\}$ and $i = 1,2,3,4$, up to translations by a lattice vector $\bm{R} = n_1\bm{a}_1 + n_2\bm{a_2}$ with $n_1,n_2 \in \mathbb{Z}$, resulting in what we refer to as a $\mathcal{C}_2$ centered primitive unit cell (see  Appendix~\ref{appsec:A_c2t_and_reality} for further discussion). We thus see that for a given lattice configuration, four BL origin gauge choices that yield such $\mathcal{C}_2$ symmetric primitive unit cell exist: Starting with any configuration, the other three choices could be generated through shifting the origin by half a primitive Bravais lattice vector in either direction---that is translations of the BL by $\pm\frac{\bm{a}_1}{2}$, $\pm\frac{\bm{a}_2}{2}$ or $\pm\frac{\bm{a}_1 + \bm{a}_2}{2}$. Such shifts correspond to an overall sign applied to either or both of the phase matrices $V(\bm{b}_1)$ and $V(\bm{b}_2)$ [a shift by $\pm\frac{\bm{a}_i}{2}$ swaps the sign of $V(\bm{b}_i)$, keeping it real]. 
Additionally, we note that the phase matrices do not depend on the geometry of the BL as only the position of the atomic orbitals relative to the Bravais lattice vectors affects the entries in Eq.~\eqref{eq:BZBC&Vmat_defn}. 


Turning to the specific case of three-band Euler lattice systems as the example we pick the unique BL origin gauge choice that gives $\text{det}\left(V(\bm{b}_1)\right) = \text{det}\left(V(\bm{b}_2)\right) = +1$ to ensure that the BZBCs~\eqref{eq:BZBC&Vmat_defn} do not amount to a discontinuous change of handedness of the eigenstate frame $\mathcal{R}(\bm{k})$, but instead amount to rotating it ($V(\bm{G}) \in SO(3)$). The reasons for this will be apparent when we discuss the atomic limit in Sec.~\ref{subsubsec:EulerConstraint}. With this, an important observation is that $V(\bm{G})$ can take four unique possibilities:
\begin{align} \label{Eq:V_choices}
    v_0 &= \text{diag}\left(1, 1, 1\right)\equiv \mathbb{1}, & v_1& = \text{diag}\left(-1, -1, 1\right) , \nonumber \\
    v_2 &= \text{diag}\left(1, -1, -1\right), & v_3& = \text{diag}\left(-1, 1, -1\right),
\end{align}
which amount to $\pi$ rotations about the axes spanning the orthonormal frame of Bloch eigenstates~\cite{Breach24PRLInterferometry}.

\begin{figure}
    \includegraphics[width=.95\linewidth]{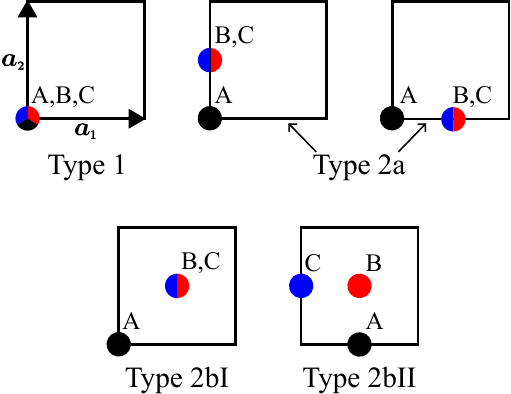}
    \caption{ Demonstration of the 5 unique possible primitive unit cells of $\mathcal{C}_2$ symmetric lattice configurations with three sublattice sites $\alpha = A,B,C$ in 2D, where generally a parallelogram unit cell is possible instead of a square one. Sublattice sites are shown as black, red or blue, with overlaps depicted by multi-colored dots. The lattice configurations are categorized by type according to their corresponding Brillouin zone boundary conditions (BZBCs) in reciprocal space as per our classification in Table~\ref{Tab:classification}. 
    Uniqueness of these possibilities means that first we disregard the $\mathcal{C}_2$ preserving gauge shifts of the unit cell origin by $\frac{\bm{a}_1}{2}$, $\frac{\bm{a}_2}{2}$ or $\frac{\bm{a}_1+\bm{a}_2}{2}$ since they correspond to the same physical lattice configuration. Accounting for these gauge choices gives four different variants for each possibility. Second, permutations of atomic site labels are neglected, which if instead are considered to be unique would lead to the 16 unique possibilities mentioned in the main text. }
    \label{Fig:UniqueUnitCells}
\end{figure}

Given that the 2D system hosts two phase matrices $V(\bm{b}_1)$ and $V(\bm{b}_2)$, we establish that there is a total of $4\times4-1 = 15$ unique lattice configurations with {\it non-trivial} BZBCs from the possibilities given in Eq.~\eqref{Eq:V_choices}
, where we excluded the \textit{trivial} BZBC case $V(\bm{b}_1) =V(\bm{b}_2)= \mathbb{1}$ which corresponds to lattice geometries where the atomic orbitals are overlapping. This is with the note that if the labeling of the three atomic sites is not physically unique, the permutations of atomic site labels are excluded, which reduces to five unique lattice configurations depicted in Fig.~\ref{Fig:UniqueUnitCells}.

We categorize all BZBC cases into four main groups based on non-triviality across the two lattice directions which can be enumerated as follows:
\begin{itemize}
    \item Type 1: $V(\bm{b}_1) = V(\bm{b}_2) = \mathbb{1}$, \qquad\qquad(trivial case)
    \item Type 2a: $V(\bm{b}_i) = \mathbb{1}, \newline V(\bm{b}_j) \in \left\{v_1,v_2,v_3 \right\} |i,j = 1,2\ , \ i \not=j$, \;\,(6/15 cases)
    \item Type 2bI: $V(\bm{b}_1), V(\bm{b}_2) \in \left\{v_1,v_2,v_3 \right\}| V(\bm{b}_1) = V(\bm{b}_2)$, 
    \qquad\qquad\qquad\qquad\qquad\qquad\quad\;\;(3/15 cases)
    \item Type 2bII: $V(\bm{b}_1), V(\bm{b}_2) \in \left\{v_1,v_2,v_3 \right\}| V(\bm{b}_1) \not= V(\bm{b}_2)$, \qquad\qquad\qquad\qquad\qquad\qquad\quad\;\;(6/15 cases)
\end{itemize}

We label the trivial BZBC as Type 1, for which the unique domain of the Bloch vectors in $\bm{k}$-space is a single BZ. However, Type 2 cases with non-trivial BZBCs demand more attention as we address in this paper in detail. We emphasize that since $V(\bm{G})^2 = \mathbb{1}$ for any combination of these non-trivial BZBCs, this classification shows that the unique domain of Bloch vectors is in fact two BZs for Type 2a and four BZs for Type 2b, which will govern the Euler invariant parity as will be discussed next.

\begin{table}
    \centering
    \centering
    \caption{Summary of the Euler class parity classification of three-band Euler systems with generally non-trivial BZBCs that occur in $\mathcal{C}_2 \mathcal{T}$ symmetric lattices in 2D with inequivalent atomic sites. The type number is 1 for fully trivial BZBCs and 2 for non-trivial ones. The further sub-types of 2 as a/b correspond to non-trivial BZBCs along one/both direction of reciprocal space respectively.}
    \label{Tab:classification}
    \begin{tabular}{>{\centering\arraybackslash}p{0.1\linewidth}>{\centering\arraybackslash}p{0.55\linewidth}>{\centering\arraybackslash}p{0.35\linewidth}}
        \hline
        \hline
        Type&  $V(\bm{b}_1),V(\bm{b}_2) \in$ 
        & Parity of $\chi_{i,i+1}(1\text{BZ})$\\
        \hline
        1&  $\left\{ \mathbb{1} \right\}$ 
        & \multirow{3}{\linewidth}{$\in 2\mathbb{Z}$ (Skyrmionic)} \\
        2a&  $\left\{ \mathbb{1} \right\}$, $\left\{ v_1,v_2,v_3 \right\}$
        & \\
        2bI&  $\left\{ v_1,v_2,v_3 \right\}$$ | V(\bm{b}_1) = V(\bm{b}_2)$ 
        & \\
        \cline{3-3}
        2bII&  $\left\{ v_1,v_2,v_3 \right\} $$| V(\bm{b}_1) \not= V(\bm{b}_2)$ 
        & $\in 2\mathbb{Z} + 1$ (Meronic) \\
        \hline
        \hline
    \end{tabular}
\end{table}

\subsubsection{Stability of Euler Topology Under Non-trivial BZBCs}\label{subsubsec:EulerConstraint}

For gapped 
Euler systems, it is important to characterize the effects of the non-trivial BZBCs on the Euler class $\chi_{2,3}$, with the $i=1$th band assumed to be the gapped one in a three-band real insulator. This can be understood by considering how the overlap of two eigenvectors in the relevant two-band subspace $O_{2,3}(\bm{k})=\braket{u_2(\bm{k})|u_{3}(\bm{k})}$ transforms upon a lattice translation by $\bm{G}$,
\begin{align} \label{Eq:EuClassBZBC}
    \!\!\! O_{2,3}(\bm{k}+\bm{G})&:= \braket{\bm{u}_2(\bm{k}+\bm{G})|\bm{u}_{3}(\bm{k}+\bm{G})} \nonumber \\
    &= s_2 s_{3} \bra{\bm{u}_2(\bm{k})} V(\bm{G})^\dag V(\bm{G}) \ket{\bm{u}_{3}(\bm{k})} \nonumber \\
    &= s_2 s_{3} O_{2,3}(\bm{k}) \nonumber\\
    &= +O_{2,3}(\bm{k}),\, \text{ for}\ s_n = 1 \;\forall n
\end{align}
where in the last line we show the overlap is indeed periodic across one BZ translations for the periodic gauge convention of the BZBCs. Clearly, Eu$_{2,3}(\bm{k})$, $\bm{A}_{2,3}(\bm{k})$, and thus $\chi_{2,3}$ transform similarly, implying that the Euler charge of nodes is not affected by translations by $\bm{G}$. This is indeed expected since the Euler class measures the relative twist between adjacent Bloch vectors, and so would be immune to an overall rotation of the set of Bloch eigenvectors that span the BZ by $V(\bm{G})$. Consequently, the BZ torus can be regarded as a closed manifold as far as these quantities are concerned and thus $\chi_{2,3}$ over a single BZ is related to the winding of $\bm{n}(\bm{k})$ as per the first line of Eq.~\eqref{Eq:EuClass_n_Winding}.

However, for non-trivial $V(\bm{G})$, since the periodic unit of the Bloch eigenvectors is $p = 2$ and $4$ BZ patches for Type 2a and 2b systems respectively, we stress that it is only over such a periodic region will the winding of $\bm{n}(\bm{k})$ in Eq.\eqref{Eq:EuClass_n_Winding} be integer quantized.
We, therefore, establish that when the system has non-trivial BZBCs, the quantization constraint on the Euler class is that it is even over a patch of $p$BZs: $\chi_{2,3}(p\text{BZs}) = p\times \chi_{2,3} \in 2\mathbb{Z}$, where we have made use of the periodicity of $\chi_{2,3}$ across the BZ. Consequently, we conclude that for systems with non-trivial $V(\bm{G})$, the quantization condition for the Euler class over a single BZ is given by:
\begin{align} \label{Eq:chi1of4BZ}
    \chi_{2,3} & = \frac{\chi_{2,3}(p\text{ BZ}) \in 2\mathbb{Z}}{p}, \quad p=2,4\nonumber \\
    & = \frac{1}{2 \pi} \int_{(\text{1 BZ})} dk^2\ \bm{n}(k) \cdot \left( \partial_{k_1}\bm{n}(k) \times \partial_{k_2} \bm{n}(k) \right) \nonumber \\
    & \in \{0,\pm1,\pm2...\}\equiv \mathbb{Z}  \quad \because \exists\ \mathcal{C}_2
\end{align}
under the $C_2$ symmetry~\footnote{$C_2$ dictates that nodes are formed in pairs and periodically across the BZs and thus $\chi_{2,3}(\text{1BZ})$ cannot take half integer values}. Notably, this shows that the Euler class over the BZ can take odd integer values in the presence of non-trivial BZBCs, which corresponds to half integer quantization of $\bm{n}(\bm{k})$ and is referred to as meronic. Moreover, since nodes are formed in oppositely-charged pairs, braiding of nodes in adjacent gaps cannot change the parity of $\chi_{2,3}$, suggesting that the parity is determined by the BZBCs of the system. 

Indeed, we find that the effect of the BZBCs on the parity of the Euler class can be understood by analyzing the Dirac strings (DSs) that arise in the atomic limit, which is the limit where tunneling amplitudes vanish and the atomic sites are decoupled. Dirac strings denote the $\pi$ Berry phase discontinuity to having a smooth gauge upon parallel transporting the Bloch eigenvectors along a non-contractable path~\cite{Ahn2019_PRX_Failure,jiang21NatPhysexperimental,Breach24PRLInterferometry}, as detected by the Zak phase \eqref{Eq:Zak_phase}~\cite{Slager24NatCommNon-Abelian}. 
Moreover, the non-Abelian nature of Euler topology results in band nodes swapping their frame charge upon crossing a DS of an adjacent gap, offering a useful perspective to braiding~\cite{Bouhon2020_NatPhys_ZrTe,jiang21NatPhysexperimental,Slager24NatCommNon-Abelian}.

To understand how Dirac strings arise in the atomic limit, we note that in this limit the Hamiltonian is diagonal with flat bands that can be set as $\epsilon_1<\epsilon_2<\epsilon_3$ for energies $\epsilon_n$, and with eigenvectors being the unit basis $\bm{u}_n(\bm{k}) = (\delta_{1n},\delta_{2n},\delta_{3n})^\intercal$ for $n = 1,2,3$. However, for lattices with non-coinciding atomic orbitals (see Type 2 in Fig.~\ref{Fig:UniqueUnitCells}), the non-trivial BZBCs amount to a sign flip of the Bloch vectors $\bm{u}_n(\bm{k})$ across the BZ for each $n^{\text{th}}$ diagonal term of $V(\bm{G})$ that is negative. Consequently, the phase matrix corresponds to non-contractible Dirac strings in the atomic limit, i.e.~an obstruction~\cite{Meron_paper,Slager24NatCommNon-Abelian} as depicted in Fig.~\ref{Fig:Classification}(a). 

We, thus, establish that the atomic limit Dirac string pattern of a system with given BZBCs along the reciprocal directions sets the possible parity of the Euler class $\chi_{2,3}$ upon braiding of nodes by tuning the system parameters, e.g.~tunneling amplitudes~\cite{Slager24NatCommNon-Abelian}, to reach a phase with well-defined $\chi_{2,3}$. This is illustrated in Fig.~\ref{Fig:Classification}(b) and (c).  Moreover, our gauge choices of det$(V(\bm{G}))= s_n = +1$ ensures that atomic limit Dirac strings correspond to a $\pi$ discontinuity for a pair of bands at a time (see Appendix~\ref{sec:BLorigin_gauge}), as is the case with Dirac strings that arise from node formation and braiding in Euler systems~\cite{Bouhon2020_NatPhys_ZrTe,jiang21NatPhysexperimental,Slager24NatCommNon-Abelian}. This underpins their natural role in determining the effect of atomic limit on the parity of the Euler class. 

Since atomic limit Dirac strings correspond to Zak phases that originate purely from the orbital positions and not the winding of the Bloch eigenvectors, such contributions are hence referred to as the \textit{orbital} Zak phases. We emphasize that these atomic limit Dirac strings persist in the system as tunnelings are continuously introduced, such that the overall measured Zak phase corresponds to both the orbital and the intrinsic band contributions stemming from band inversions~\cite{Slager24NatCommNon-Abelian}. Consequently, a given phase of an Euler system with non-trivial BZBCs will have to be compared with the reference atomic limit to distinguish the orbital and band contributions of Zak phase measurements (see also Ref.~\cite{Meron_paper}).

\subsection{Parity Classification of 2D Euler Phases for Three Bands \& Beyond}
\label{subSec:Eu_Parity_Classification}

Starting from the reference point of the atomic limit and based on the the possible DS configurations therein, one can now establish the Euler class parity classification which is summarized compactly in Table.~\ref{Tab:classification}. We enumerate these for the possible cases introduced previously and in Fig.~\ref{Fig:UniqueUnitCells} as follows:

\textit{Type 1:} The atomic limit BZ has no Dirac strings and the periodic unit is one BZ, so the Euler phases exhibited are even, i.e.~skyrmionic, ($\chi_{i,i+1} \in 2\mathbb{Z}$).

\begin{figure}    \includegraphics[width=.95\linewidth]{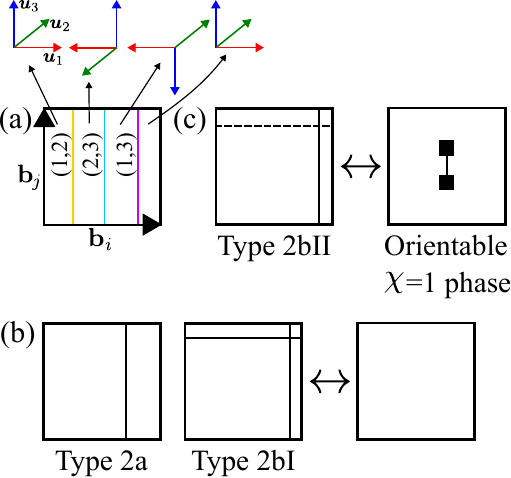}
    \caption{ a) Illustration of the atomic limit Dirac strings within one BZ that follow from the non-trivial BZBCs shown in Eqs.~\eqref{eq:BZBC&Vmat_defn} and \eqref{Eq:V_choices}. The yellow, cyan and purple lines respectively show  non-contractible DS loops arising from $V(\bm{b}_i)= v_1,v_2,v_3$ for $i=1,2$, corresponding to different gaps between the bands specified next to each string. The orthonormal eigenstate frame shown above gets rotated upon crossing a DS [e.g.~yellow DS] due to the sign flips of the relevant eigenstates [($u_1,u_2$)]. b) Black lines correspond to a DS of any of the three possible types (colored in (a)). Systems classified as Type 2a and 2bI in Table~\ref{Tab:classification} have atomic limit DS configurations that are continuously connected to a fully orientable flat band state. c) Dashed lines indicate a DS of any type different from that of the solid line. The only fully orientable state that is continuously connected to the atomic limit Dirac string configuration of Type 2bII is that with an odd (meronic) Euler class $\chi = 2\mathbb{Z}+1$. 
    }
    \label{Fig:Classification}
\end{figure}

\textit{Type 2a:} The atomic limit has a single non-contractible Dirac string as shown in Fig.~\ref{Fig:Classification}(b), which can be continuously removed via creation of an appropriate pair of band nodes and annihilating them across the BZ, resulting in zero Euler class for all bands. Since nodes are formed in pairs with opposite Euler charges, the only possibility is skyrmionic phases ($\chi_{i,i+1} \in 2\mathbb{Z}$).

\textit{Type 2bI:} This is type 2b with the additional condition $V(\bm{b}_1) = V(\bm{b}_2)$, which corresponds to Dirac strings of the same type (i.e.~between the same bands) across the two reciprocal vector directions (see Fig.~\ref{Fig:Classification}(b)). Since the Euler charge of nodes is not affected upon crossing Dirac strings in the same two-band subspace, again the atomic limit of these systems is continuously connected to a state with zero Euler class in the same way as in Type 2a. Hence, only skyrmionic phases ($\chi_{i,i+1} \in 2\mathbb{Z}$) are possible. 

\textit{Type 2bII:} That is type 2b but with different types of Dirac strings along the reciprocal lattice directions $V(\bm{b}_1) \not= V(\bm{b}_2)$. Attempting to remove the Dirac strings by continuous creation and braiding of nodes gives a single-gap state with a pair of nodes with similar Euler charges such that $\chi_{i,i+1}= 1$ in the two-band subspace. There is no continuous connection to a fully gapped and orientable state as shown in Fig.~\ref{Fig:Classification}(c). 
Consequently, the only possibility is meronic phases ($\chi_{i,i+1} \in 2\mathbb{Z}+1$).

We note that while our results above count all possible cases for three-band real Hamiltonians, for systems with more than three bands our characterization applies in a similar way. 
The crucial point is that the parity of the resulting Euler invariants will be constrained by the atomic orbital positions, i.e.~Wannier obstructions. One should start with identifying all possible lattice configurations as in Fig.~\ref{Fig:UniqueUnitCells} where we consider $\mathcal{C}_2$-centered sites and BL origin that results in with $\det V(\bm{b}_1) = \det V(\bm{b}_2)= +1$ as much as possible, establish the corresponding atomic limit Dirac strings for each configuration, and consider which Euler invariants that are orientable within their subspaces can be obtained from there upon inducing band nodes and braiding them. The only way to attain odd Euler invariants within a sub-set of three adjacent bands again requires lattice geometries in which the atomic limit Dirac strings along $\bm{b}_1$ and $\bm{b}_2$ are different and lie within adjacent gaps (see Fig.~\ref{Fig:Classification}(c)). Then, the interplay between different subspaces would need to be considered to obtain the overall parity of the invariants. Moreover, while the Euler class is defined only in two dimensions, one can define it on 2D slices of higher dimensional spaces~\cite{Jankowski24PRBHopf,Adrien20PRB_Geometric,Bouhon23Arxivquantum,Bouhon22ArxivMulti}, in which case our procedure can be applied by accounting for the BZBCs along cuts.

The main takeaway from this classification is that non-trivial phase matrix $V(\bm{G})$ BZBCs amount to the presence of orbital Zak phases and associated Dirac strings in the BZ, which only affects the Euler topology by dictating the possible parity of the Euler invariant 
that can be accessed by forming, braiding, and annihilating singularities. Beyond the determination of the parity of the multi-gap invariant due to such atomic limit Dirac strings, the Euler topology of Type 2 systems is unaffected by the non-trivial BZBCs and is rather analogous to that of Type 1 that are single-BZ periodic systems in that the invariant, node charges and the eigenspectrum (including nodes) are BZ periodic. Moreover, we also conclude that meronic Euler phases with an odd invariant 
are only possible for systems of Type 2bII, that is, only when the BZBCs are non-trivial and anisotropic across both directions of momentum-space. This odd Euler class is a direct consequence of the obstruction that arise from the non-trivial BZBCs, and indeed experimentally measurable. Consequently, meronic single-gap phases exhibit half integer quantized winding per BZ of $\bm{n}(\bm{k})$, the Bloch eigenvector of the third (gapped) band, in agreement with Eq.~\eqref{Eq:chi1of4BZ}. 

In the remainder of this manuscript, we address this question of observable signatures of non-trivial BZBCs that can appear dynamically. In particular, we extend the analysis of quench dynamics that have been established previously considering only the single-BZ periodic Type 1 phases~\cite{Even_Quench} to the cases of Type 2 phases involving topological obstructions. As such, we obtain smoking-gun signatures that capture the physics of the non-trivial $V(\bm{G})$ BZBCs, further hammering in the gauge invariant physical consequences of the $V(\bm{G})$ rotations of Bloch vectors at the BZ boundary.

\section{Quench Dynamics With Non-trivial BZBCs}\label{sec:QuenchDynamics}

Quantum quenches have been established in recent years as a powerful tool to study out-of-equilibrium dynamics of topological systems, unearthing new kinds of topologically protected responses and offering versatile probes for experiments~\cite{Wang17PRLChern, Tarnowski19_NatCom,Unal2019PRR,Sun2018_PRL_Quench,yang2018_PRB_Quench}. These ideas have revealed a monopole--anti-monopole structure in momentum-time trajectories of Euler insulators when only trivial boundary conditions are considered~\cite{Even_Quench,EvenQuenchExper}, i.e.~for Type 1 in our classification. We here start with a brief summary of these results which essentially involve dynamical linking patterns manifesting through Hopf maps, before analytically demonstrating whether and how one can apply quench dynamics in the presence of non-trivial BZBCs for both skyrmionic (Type 2a \& Type 2bI) and meronic (Type 2bII) Euler phases. 
We further provide numerical simulations in kagome lattice (shown in Fig.~\ref{Fig:MeronModel}(a)) for the more informative meronic phases when contrasting with the Type 1 case~\cite{Even_Quench} to confirm our procedure; skyrmionic Type 2 phases are indeed more similar to the Type 1 case as we demonstrate below. We illustrate the resulting monopole--anti-monopole physics of odd Euler phases along with how it is affected across the BZ by the phase matrices $V(\bm{G})$.

\subsection{Review of Quench Dynamics of Type 1 Euler insulators}
The topology of a Bloch Hamiltonian can manifest as links in momentum-time space, imprinted via the Hopf map upon quenching by said Hamiltonian. 
This idea was first employed for two-band Chern insulators, both for in- and out-of-equilibrium topological phases~\cite{Wang17PRLChern, Unal2019PRR, Tarnowski19_NatCom,Chen2020_PRA_linking,Yi2019_Arxiv_Hopf,hu2020_PRL_TopInv}. A two-band Chern insulator has a Hamiltonian 
that generally takes the form $H_C (\bm{k}) = \bm{d}(\bm{k})\cdot \bm{\sigma} + d_0 \mathbb{1}$, for Pauli matrices $\bm{\sigma}$ and the Hamiltonian vector $\bm{d}(\bm{k})$. The Chern number $C$ is a quantized integer that corresponds to the number of times the corresponding unit vector $\hat{\bm{d}}(\bm{k}) = \frac{\bm{d}(\bm{k})}{|\bm{d}(\bm{k})|}$ covers the Bloch sphere \cite{Pont1941,Hasan2010_Rev_colloquium},
\begin{equation} \label{Eq:Chern_wind}
    C = \frac{1}{4\pi} \int_{\text{BZ}} d^2k \ \hat{\bm{d}} \cdot \left(\partial_{k_x} \hat{\bm{d}} \times \partial_{k_y}\hat{\bm{d}}  \right).
\end{equation}

Quenching an initial trivial state $\bm{\psi}_0(\bm{k})$ by the target Hamiltonian $H_C(\bm{k})$ and then projecting the time-evolving state to the Bloch sphere gives a time periodic Bloch vector $\bm{q}(\bm{k},t)\subset S^2$, such that the ($k_x,k_y,t)$ paramater space forms a three torus~\footnote{Time evolution forms a circle $S^1$, which is completed after $2\pi$ rotation by the state upon considering spectrally flattening the Hamiltonian $H_C (\bm{k})$, which does not change the Chern number. This ensures that the state returns to itself at the same time for all $\bm{k}$, while the Hopf-linking invariant is indeed stable even when relaxing this conditions as observed in experiment~\cite{Tarnowski19_NatCom}}. 
Since the weak invariants are vanishing, we have $T^3 \cong S^3$ which establishes a (first) Hopf map $S^3 \to S^2$ for $\bm{q}(\bm{k},t)$~\cite{Wang17PRLChern,Unal2019PRR}.  
Crucially, given the Hopf map, the linking number $\mathcal{L}$ of the inverse images in $T^3$ of any two points $\bm{q}_i\in S^2$ equals the Hopf invariant $\mathcal{H}$, which in turn can be shown to be precisely given by the difference in Chern numbers of the initial state $\bm{\psi}_0$ and the quenching Hamiltonian $H_C(\bm{k})$, a result that has been established both theoretically and experimentally~\cite{Wang17PRLChern,Tarnowski19_NatCom}. 

\begin{figure}
    \includegraphics[width = 0.95 \columnwidth]{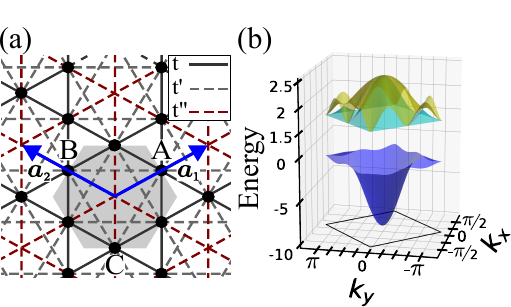}
    \caption{ a) The Kagome lattice with tunnelings for nearest $(t)$, next-nearest $(t')$, and third-next-nearest $(t'')$ neighbours. 
    b) Dispersion of the meronic phase of the system which has $\chi_{23}=1$ with the parameters $t=t'=-1$ and $t''=-0.8$.}
    \label{Fig:MeronModel}
\end{figure}

While such an application of the Hopf fiber and linkings in momentum-time trajectories evidently fails beyond two-level systems, an analogous map has been developed for three- and four-band Euler insulators by taking advantage of the reality condition under $\mathcal{C}_2\mathcal{T}$ symmetry. Notably, this was possible by considering Bloch Hamiltonians $H(\bm{k})$ which are periodic over the BZ with no Wannier obstructions~\cite{Jankowski24PRBHopf,Even_Quench,Jankowski25PRBoptical}. 
For the Euler subspace $\chi_{2,3}$ (c.f.~Fig.~\ref{Fig:MeronModel}(b)), the procedure~\cite{Even_Quench} follows in parallel by employing a spectrally-flattened Euler Hamiltonian which can be expressed in terms of the gapped eigenstate $\bm{n}(\bm{k})\equiv \bm{u}_1(\bm{k})$ as~\cite{Bouhon2020_NatPhys_ZrTe},
\begin{equation} \label{Eq:H_flat}
    H_{\text{flat}} = 2 \bm{n}(\bm{k}) \cdot  \bm{n}(\bm{k})^\intercal - \mathbb{1}_3,
\end{equation}
where the two-band (Euler) subspace becomes degenerate and harbors the non-Abelian band nodes. Thus, similar to the Chern case, the Euler Hamiltonian is now written in terms of a unit vector $\bm{n}(\bm{k})$ that spans the two-sphere and for which the Euler class is given by its winding through Eq.~\eqref{Eq:EuClass_n_Winding}. We note the factor of two difference between Eq.~\eqref{Eq:Chern_wind} and Eq.~\eqref{Eq:EuClass_n_Winding}, which will bring an important distinction in the Euler case.

Starting with an initial trivial state $\bm{\Psi}(\bm{k},t=0) = \bm{\Psi}_0 \  \forall \bm{k}$ and quenching by the Hamiltonian $H_{\text{flat}}$ \eqref{Eq:H_flat} yields a three-torus $(k_x,k_y,t)$ parameter space $T^3$. Then, projecting the evolved state 
to a time periodic Bloch vector $\bm{p}(\bm{k},t)$ defines a Hopf map  as $\bm{p}(\bm{k},t): T^3\cong S^3 \to S^2$. This projection to the two-sphere is established by the non-trivial relation $\bm{p}(\bm{k},t) = \bra{\bm{\Psi}^\dag(\bm{k},t)} {\boldsymbol{\mu}} \ket{\bm{\Psi}(\bm{k},t)}$, which is developed by utilizing a quaternion description of the Hopf map to obtain the $\bm{\mu}$ matrices that address the three-level wave function and play the role of the regular Pauli matrices (for details see Appendix C of~\cite{Even_Quench}). The linking signatures 
in the momentum-time parameter space thus reveal the non-Abelian braiding of the band nodes in the Hamiltonian $H_{\text{flat}}$, allowing for their experimental observation~\cite{EvenQuenchExper}.

The relevant unit vector governing the inverse image linkings is shown to be $\bm{a}(\bm{k}) \coloneq H_{\text{flat}} \bm{\Psi}_0$~\cite{Even_Quench}, which is a $\pi$ rotation of $\bm{\Psi}_0$ around $\bm{n}(\bm{k})$, implying $\bm{a}(\bm{k})$ wraps $S^2$ twice as much as $\bm{n}(\bm{k})$ within the BZ, relating to the factor of two difference highlighted above. Consequently, $\bm{a}(\bm{k})$ is shown to host a monopole--anti-monopole structure through the orientation of its winding relative to that of $\bm{n}(\bm{k})$. Specifically, within the $\bm{k}$-space region where $\bm{n}(\bm{k})$ lies in the $+\bm{\Psi}_0$($-\bm{\Psi}_0$) hemisphere (see Fig.~\ref{Fig:n_k_Cover_Meron}(a) and (c) for an illustration respectively), $\bm{a}(\bm{k})$ winds (unwinds) the two-sphere defining a monopole (anti-monopole). This behavior manifests itself as linkings in $T^3$ with a $+$ and $-$ sign. We refer to these opposite signs of linkings as linking polarity and emphasize that they occur in different BZ patches corresponding to the opposite hemispheres where the winding and unwinding occurs respectively, which together reflect the double winding of $\bm{a}(\bm{k})$ relative to $\bm{n}(\bm{k})$. 

The winding of $\bm{a}(\bm{k})$ over a BZ patch $\mathcal{D}_{+/-}$ is associated with linking number $\mathcal{L}$ and corresponding Hopf invariant $\mathcal{H}$, which has been shown to be written as~\cite{Even_Quench} 
\begin{equation} \label{Eq:H_aLink}
    \mathcal{L}_\mathcal{D} = \mathcal{H}_\mathcal{D} =  \frac{1}{4\pi} \int_{\mathcal{D}} dk^2 \bm{a}(\bm{k}) \cdot \left( \partial_{k_1}\bm{a}(\bm{k}) \times \partial_{k_2}\bm{a}(\bm{k}) \right) .
\end{equation}
Here, the value of the invariant for a patch $\alpha=\mathcal{D}_{+} (\mathcal{D}_{-})$ corresponding to monopole (anti--monopole) is referred to as the monopole charge $\mathcal{H}_{\alpha} =  +1 (-1)$. Since the winding of $\bm{a}(\bm{k})$ is related to twice that of $\bm{n}(\bm{k})$ (and hence is equal in magnitude to Eu$_{2,3}$($\bm{k}$)) up to an orientation, we have that $\chi_{2,3} = \sum_{\alpha \in \text{BZ}} |\mathcal{H}_{\alpha}|$ (where we implicitly mean the Euler invariant over a single BZ). Note that $\mathcal{H}$ and thus the Hopf physics is independent of the choice of $\bm{\Psi}_0$, albeit the later being affected by a change in the Hopf parametrization and pattern of inverse images~\cite{Even_Quench}. Likewise, these considerations still hold for quenching Hamiltonians with dispersive bands~\cite{Tarnowski19_NatCom} 
emphasizing that the signitures are hinged on the Euler topology. 

We end by noting a new insight on the dynamics of Euler systems. For two-level Chern models, the time-evolving Bloch vector $\bm{q}(\bm{k},t)$ precesses about the $\hat{\bm{d}}(\bm{k})$ vector following a quench for a given coordinate $\bm{k}$, which explains why regions in momentum space in which $\hat{\bm{d}}(\bm{k})$ covers the full $S^2$ Bloch sphere $\bm{q}(\bm{k},t)$ defines a Hopf map~\cite{Wang17PRLChern,Tarnowski19_NatCom}. We find that an analogous precession picture can be adopted also for the three-level Euler case where for a given $\bm{k}$ coordinate, $\bm{p}(\bm{k},t)$ precesses about the projected $\Tilde{\bm{n}}(\bm{k}) = \bra{\bm{n}^\dag} {\boldsymbol{\mu}} \ket{\bm{n}}$ vector which winds the Bloch sphere twice as much as the original eigenvector of the gapped band $\bm{n}(\bm{k})$ as we confirm numerically. 
Hence $\bm{p}(\bm{k},t)$ defines a Hopf map over each momentum-space patch $\mathcal{D}_{+/-}$ where $\bm{n}(\bm{k})$ covers half of the Bloch sphere $S^2$. This viewpoint underpins the fundamental difference of the Hopf maps in Euler class with the Chern case and provides insight for addressing the non-periodic BZBCs. From this we conclude that in the flat Hamiltonian limit \eqref{Eq:H_flat}, the quench dynamics of a three-level Euler insulator with $\chi_{i,i+1}= m$ are equivalent to that of an effective two-level Chern system with $\hat{\bm{d}}(\bm{k}) = \Tilde{\bm{n}}(\bm{k}) $ with $C=m$, a result that nicely resonates with the correspondence between the two topologies~\cite{Bouhon2020_NatPhys_ZrTe}. 

\subsection{Quench Dynamics of Type 2 Euler Phases and Signatures of Obstruction} \label{sec:Quench_Results}

We now turn to the obstruction effects in Type 2 Euler phases with non-trivial BZBCs. The central idea we conclude is that precisely because the stability of Euler topology across the BZ is unaffected by the BZBCs (i.e.~it is invariant over the BZ as established in Sec.~\ref{Sec:OddEu}), as long as a consistent gauge is picked for the eigenvectors 
as we showed in Eq.~\eqref{Eq:EuClassBZBC}, the quench dynamics of Type 2 cases can be constructed building on the protocol for (skyrmionic) Type 1 phases summarized above~\cite{Even_Quench}. 
The main highlight is that we demonstrate meronic Type 2 phases exhibit an odd number of links  while skyrmionic ones manifest an even number of them per BZ. Moreover, the non-trivial BZBCs must be factored in correctly to be able to observe these odd linking numbers which govern how inverse images relate between adjacent BZs. These form a direct physical signature for odd Euler physics and obstruction.

We find that for Type 2 systems, distinctively, the phase matrix $V(\bm{G})$ transformation of $\bm{n}(\bm{k})$ upon translations by a reciprocal vector $\bm{G}$ is carried forward to the Hopf linking patterns as new transformation relations across the BZ that depend on the form of $V(\bm{G})$ and the choice of initial state $\Psi_0$. This in some cases also amount to polarity swap of the linking patterns between adjacent BZs, enforced by obstruction. More specifically, these relations determine how the Bloch vector $\bm{p}(\bm{k},t)$ is affected upon translations by a reciprocal lattice vector $\bm{G}$. Considering an initial state $\Psi_0 = \hat{\bm{x}} = (1,0,0)^\intercal$, we derive them to be as follows (see Appendix~\ref{AppSec:Quench} for details):
\begin{align}\label{Eq:p_relations}
    \bm{p}(\bm{k}+\bm{G}, t) = 
    \begin{cases}
        \bm{p}(\bm{k}, t) & \text{if $V(\bm{G})=v_0$}, \\ \\
        -v_2 \cdot \Tilde{\bm{p}}(\bm{k}, t) & \text{if $V(\bm{G})=v_1$}, \\ \\
         v_1 \cdot \bm{p}(\bm{k}, t) & \text{if $V(\bm{G})=v_2$}, \\ \\
         -v_3 \cdot \Tilde{\bm{p}}(\bm{k}, t) & \text{if $V(\bm{G})=v_3$},
    \end{cases}
\end{align}
where $\Tilde{\bm{p}}(\bm{k}, t)$ is the same Bloch state $\bm{p}(\bm{k}, t)$ but with $\bm{a}(\bm{k}) \to -\bm{a}(\bm{k})$ in it is expression, and thus corresponds to swapped polarity for the linking inverse images. Indeed, since a monopole (anti-monopole) corresponds to $\bm{n}(\bm{k})$ pointing along the $\Psi_0 = +\hat{\bm{x}}$ ($-\Psi_0 = -\hat{\bm{x}}$) hemisphere (c.f.~Fig.~\ref{Fig:n_k_Cover_Meron}(a) and (c) respectively), $V(\bm{G}) = v_1$ and $v_3$ correspond to a swap of polarity as they change the sign of the $\hat{\bm{x}}$-component of the gapped eigenvector $\bm{n}(\bm{k})$. 
In general, a Bloch vector $\bm{p}(\bm{k}, t)$ at a given $\bm{k}$ in a reference BZ would transform into $\bm{p}(\bm{k} + \bm{G}, t)$ in an adjacent BZ. In other words, if a set of points $(\bm{k},t)$ correspond to the inverse image of $\bm{p}_i$, in the adjacent BZ these points $(\bm{k}+\bm{G},t)$ map to a Bloch vector  given by the appropriate transformation of $\bm{p}_i$ as determined by the relations~\eqref{Eq:p_relations}. 

Furthermore, we highlight that the difference compared to quench dynamics of Type 1 systems is that in the Type 2 case the periodic unit of the momentum-time space three-torus $T^3$ for linkings is no longer a single BZ in the $\bm{k}$-space, but instead is a region comprising 2 or 4 BZs for Type 2a and Type 2b systems respectively. These results identify how the non-periodic BZBCs materialize in a Type 2 Euler system with non-trivial invariant $\chi_{2,3}$~\footnote{Note that here we inherently assume the system is orientable for the Euler class to be well defined as a BZ periodic invariant, see Sec.~\ref{Sec:conclusion} for further discussion.} by analyzing its dynamics upon quenching a trivial initial state $\Psi_0$ and examining the resultant Hopf linking pattern in general in a four-BZ patch in the parameter space $(k_x,k_y,t)$. Comparing how linking patterns transform across adjacent BZs with the standard transformation relations~\eqref{Eq:p_relations} for the choice $\Psi_0 = \hat{\bm{x}}$ directly reveals the BZBC effects, where transformation relations for other choices can be established similarly as discussed in Appendix~\ref{AppSec:Quench}. As a result, such a scheme would allow experimental verification of the Euler class parity classification in Tab.~\ref{Tab:classification} with the odd Euler physics only arising in the presences of anisotropic non-trivial BZBCs.

\subsection{Numerical Results for a Meronic Euler Phase}

We now numerically demonstrate how these boundary obstruction effects emerge in a model kagome lattice by using a meronic Euler phase with $\mathcal{C}_2\mathcal{T}$ symmetry (where symmetries hold individually and combined)~\cite{Meron_paper}. Three atomic sites per unit cell are connected by hopping amplitudes $t,t',t''$ corresponding to nearest, next-nearest and third-next-nearest neighbours respectively, with primitive lattice vectors $\bm{a}_1$ and $\bm{a}_2$ positioned as demonstrated in Fig.~\ref{Fig:MeronModel}(a). The tight-binding Hamiltonian in the site basis is written as,
\begin{align}\label{Eq:MeronH}
 \!\!H(\bm{k}) \!=& \ 2t \sum_{\alpha \not = \beta } \text{cos}(\bm{k}\cdot \boldsymbol{\delta}_{\alpha \beta}) c^\dagger_\alpha c_\beta +  2t^\prime \sum_{\alpha \not = \beta } \text{cos}(\bm{k}\cdot \boldsymbol{\delta}^\prime_{\alpha \beta}) c^\dagger_\alpha c_\beta \nonumber \\
    & + 2t'' \sum_{\alpha} \text{cos}(\bm{k}\cdot \boldsymbol{\delta}''_{\alpha \alpha}) c^\dagger_\alpha c_\alpha  + \sum_{\alpha } \Delta_\alpha c^\dagger_\alpha c_\alpha,
\end{align}
where $c^{(\dagger)}_\alpha$ is an annihilation (creation) operator at the atomic site $\alpha$ and the $\bm{\delta}_{\alpha \beta}/\bm{\delta}^\prime_{\alpha \beta}/\bm{\delta}^{\prime \prime}_{\alpha \beta}$ are the nearest/next-nearest/third-next-nearest neighbour vectors between sites $\alpha,\beta \in\left\{ A,B,C\right\}$ which are given in Appendix~\ref{AppSec:MeronH} explicitly. Specifically, we consider the phase with $t=t' = -1$, $t'' = -0.8$ and sublattice offsets $\Delta_{\alpha}=0 \ \forall \alpha$ which features a single gap and a two-band sub-space with $\chi_{23}(\text{1 BZ}) = 1$, as shown in Fig.~\ref{Fig:MeronModel}(b). For this system, we find that $V(\bm{b}_1) = v_3$ and $V(\bm{b}_2) = v_2$ as visible from the sub-lattice configuration in Fig.~\ref{Fig:MeronModel}(a). The Hamiltonian can be spectrally flattened by considering the gapped lowest eigenstate $\bm{n}(\bm{k}) \equiv \bm{u}_1 = \bm{u}_2 \times \bm{u}_3$ via Eq.~\eqref{Eq:H_flat}.

\begin{figure}
    \includegraphics[width = .98 \columnwidth]{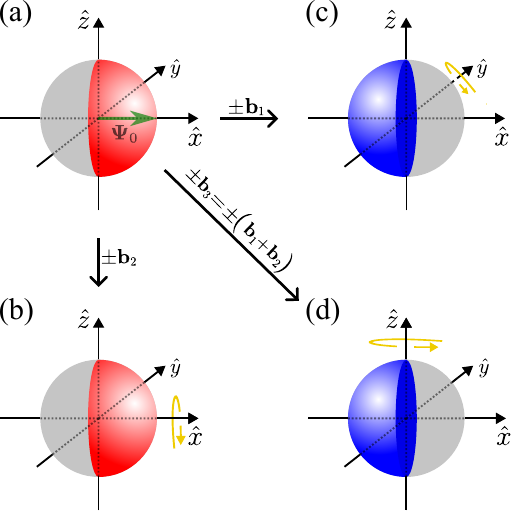}
    \caption{ Illustration of a) the coverage on the Bloch sphere by the $\bm{n}(\bm{k}) \equiv \bm{u}_1(\bm{k})$ eigenvector of the meronic Euler Hamiltonian of Eq.~\ref{Eq:MeronH}, along with b), c), and d)  how it is transformed respectively upon translations by a reciprocal lattice vector $\bm{G} =\bm{b}_1,\bm{b}_2,\bm{b}_3 $ through the $V(\bm{G})$ matrices as rotations about the Bloch sphere axes. The initial state vector $\bm{\Psi}_0 = +\hat{\bm{x}}$ for the quench dynamics is shown with a green arrow in (a), from which we see that $
    \bm{n}(\bm{k})$ is covering the $+\bm{\Psi}_0$ hemisphere. For the Kagome geometry given in Fig.~\ref{Fig:MeronModel}, $V(\bm{b}_1) = v_3$ and $V(\bm{b}_2) = v_2$. Hence, while a transformation along the first reciprocal direction rotates $\bm{n}(\bm{k})$ to the opposite hemisphere (c) and (d), it remains in the same hemisphere upon moving along $\bm{b}_2$ (b).}
    \label{Fig:n_k_Cover_Meron}
\end{figure}

Considering the quench dynamics of the given model using an initial trivial state $\bm{\Psi}_0(\bk)= (1,0,0)^\intercal$, we find that the half-sphere coverage of $\bm{n}(\bm{k})$ over a single BZ dictates the signatures as a result of the non-periodicity of the system across the BZ. We demonstrate in Fig.~\ref{Fig:n_k_Cover_Meron}(a) how this coverage spans precisely the $+\hat{\bm{x}} = +\bm{\Psi}_0 $ hemisphere, which underpins a Hopf linking monopole, in the reference BZ and indeed is half quantized following Eq.~\eqref{Eq:chi1of4BZ}. 

Since $\bm{a}(\bm{k})$ covers the Bloch sphere twice as much as $\bm{n}(\bm{k})$ as established analytically, we obtain linking number $\mathcal{L} = \mathcal{H} = +1$ also numerically for the reference BZ considered in Fig.~\ref{Fig:n_k_Cover_Meron}(a) within which $\bm{n}(\bm{k})$ manifests a monopole. To illustrate the effects of the non-trivial phase matrices (BZBCs) on the Bloch vector, we then move to the adjacent BZs. In Fig.~\ref{Fig:n_k_Cover_Meron}(b), we see that lattice translations by $\bm{b}_2$ rotate $\bm{n}(\bm{k})$ by $V(\bm{b}_2) = v_2$ which maintains the Hopf polarity. 
However, Figs.~\ref{Fig:n_k_Cover_Meron}(c) and (d) respectively show how translations by $\bm{b}_1$ or $\bm{b}_3 \coloneq \bm{b}_1 + \bm{b}_2$ rotate $\bm{n}(\bm{k})$ by $V(\bm{b}_1) = v_3$  or $V(\bm{b}_3) \coloneq V(\bm{b}_1)V(\bm{b}_2) = v_1$, thereby changing the Hopf polarity which results in a linking number $\mathcal{L}= -1$ for these translated BZs. Indeed, using~\eqref{Eq:p_relations}, we find that the corresponding evolved Bloch vector transforms upon translations in reciprocal space through $\bm{p}(\bm{k}+\bm{b}_1,t) = -v_3\cdot \Tilde{\bm{p}}(\bm{k}+\bm{b}_1,t) $ and $\bm{p}(\bm{k}+\bm{b}_2,t) = v_1\cdot \bm{p}(\bm{k}+\bm{b}_1,t)$, which predicts the presence or absence of these polarity swaps. Likewise, upon transforming $\bm{a}(\bm{k}) \to \bm{a}(\bm{k}+\bm{G})$ and utilizing the BZBC $\bm{n}({\bm{k}+\bm{G}})= V(\bm{G}) \bm{n}(\bm{k})$ in Eq.~\eqref{Eq:H_aLink}, one also attains an overall positive or minus sign in the winding integral for $\bm{G}= \bm{b}_2$ and $\bm{G}=\bm{b}_1,\bm{b}_3$ respectively, further affirming the aforementioned sign swaps of $\mathcal{L}$ for adjacent BZs as direct consequences of the obstruction.

\begin{figure}
    \includegraphics[width = 1 \columnwidth]{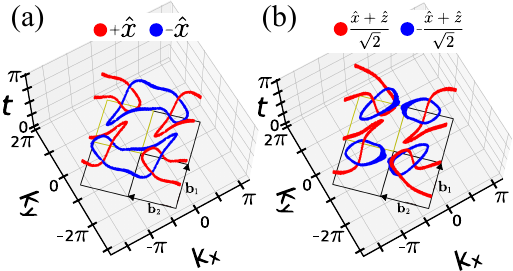}
    \caption{The linking invariants for two points on $S^2$, $\bm{p}_1$ (in blue) and $\bm{p}_2$ (in red) as specified in the legends. a) Inverse images of $\bm{p}_{1,2} = \pm \hat{\bm{x}} = (1,0,0)^\intercal$ are shown over four BZs which is the periodic unit for the linking and shown as black rectangles with the reference BZ highlighted in yellow. 
   The unusual linking pattern manifests as a chain-like structure, coupling across BZs, with one linking per BZ signaling the meronic phase. Moreover, because $\bm{p}_1 = v_1\cdot \bm{p}_2 $ and $\bm{p}_{1,2}= -v_3\cdot \bm{p}_{1,2}$, we observe that colors swap along $\bm{b}_2$, and along $\bm{b}_1$ linking is swapped to anti-linking with the same pattern (c.f.~Eq.~\eqref{Eq:p_relations}) 
   b) Inverse images for $\bm{p}_{1,2} =\pm \frac{1}{\sqrt{2}} (1,0,1) = \pm \frac{\hat{\bm{x}} + \hat{\bm{z}}}{\sqrt{2}} $ again with single linking per BZ. Because $\bm{p}_1 \not = \text{diag}\left( -1,-1,1\right) \cdot \hat{\bm{x}}_2 $ but $\bm{p}_{1,2}= \text{diag}\left( 1,-1,1\right)\cdot \bm{p}_{1,2} $, we observe a different linking pattern along $\bm{b}_2$ but an identical one, albeit with linking becoming anti-linking, along $\bm{b}_1$.}
    \label{Fig:InvImgs}
\end{figure}

These analytical characterizations in turn also reflect on the linking patterns and quench dynamics. 
We corroborate these numerically by analyzing the inverse images of the projected evolved state vector $\bm{p}(\bm{k},t) \in S^2$ for a pair of points $\bm{p}_1$ and $\bm{p}_2$ on the Bloch sphere, following the quench. 
We demonstrate the inverse images in $T^3$ parameter space ($k_x,k_y,t$) in Fig.~\ref{Fig:InvImgs} for two choices of $(\bm{p}_1,\bm{p}_2)$ pairs. Interestingly, we observe a chain-like linking manifesting across the BZs and conjoined through the non-trivial BZBCs, with one link per BZ.

As for determining the effect of the phase matrices $V(\bm{G})$ through the transformation patterns of the inverse images across the BZ boundary, we consider how choices of $(\bm{p}_{1}$,$\bm{p}_{2})$ inverse image pairs are transformed across the BZ, as dictated by the relations in Eq.~\eqref{Eq:p_relations}. In particular for the given model, since for translations by $\bm{G}=\bm{b}_2$ we have $\bm{p}(\bm{k}+\bm{b}_2, t) = v_1 \cdot \bm{p}(\bm{k}, t)$, the inverse image of a point $\bm{p}_i$ becomes that of $\bm{p}_j = v_1 \cdot \bm{p}_i$. Consequently, for $\bm{p}_{1,2} = \pm \hat{\bm{x}}$ depicted in Fig.~\ref{Fig:InvImgs}(a) we establish that the inverse images of these points swap upon a shift by $\bm{b}_2$, as confirmed visually by the color swaps of the inverse image links across the BZ along this direction. This indeed forms the bases of the chain linking. By contrast, for translations by $\bm{G}=\bm{b}_1$, the relation $\bm{p}(\bm{k}+\bm{b}_1, t) = -v_3 \cdot \Tilde{\bm{p}}(\bm{k}, t)$ implies that the linking of an inverse image of a reference point $\bm{p}_i$ changes to linking with opposite polarity (i.e.~opposite sign of $\mathcal{L}$) of the point $\Tilde{\bm{p}}_i = -v_3 \bm{p}_i$. Thus, for a choice of $({\bm{p}_1,\bm{p}_2})$ pairs that satisfy $\Tilde{\bm{p}}_{i} = \bm{p}_{i}$ for $i=1,2$, a translation by $\bm{b}_1$ swaps the polarity of the inverse image links. This is exactly the case for $\bm{p}_{1,2} = \pm \frac{\hat{\bm{x}}+\hat{\bm{z}}}{2}$ that we demonstrate in Fig.~\ref{Fig:InvImgs}(b). 
As such, we determine the specific transformation relations of inverse images along each reciprocal direction, from which the BZBCs phase matrices can be deduced correctly as $V(\bm{b}_1) = v_3$ and $V(\bm{b}_2) = v_2$. This verifies the physical consequences of the phase matrix BZBCs and that they are not mere gauge artifacts.


\section{Discussion and Conclusion} \label{Sec:conclusion}
We now discuss some implications of our characterization and results. We have here considered Euler insulating phases within the two-band subspace that are fully gapped from other bands and orientable, which guarantees a non-zero Euler invariant via Eq.~\eqref{Eq:chi1of4BZ}. The crucial idea of our work is then that the winding of the gapped Bloch eigenvector $\bm{n}(\bm{k})$ is quantized on $S^2$ for such a three-band system, be it meronic or skyrmionic, which in turn governs the quench Hopf signatures laying bare the BZBCs. However, we note that this quantization breaks down for non-orientable Euler phases which have a Dirac string in the adjacent gap. 

In the convention we follow with the Euler subspace $\chi_{2,3}$, this non-orientable case amounts to having a Dirac string between bands $i=1,2$. While the Euler class $\chi_{2,3}$~\eqref{Eq:FullEuClass} across the BZ is ill-defined due to the Dirac string discontinuity in the second band $\bm{u}_2$, it can still be calculated over a momentum-space patch, which can be treated as the entire BZ by gauging out the $(1,2)$ Dirac string~\footnote{This amounts to picking the gauge with $s_2=-1$ (while det$(V(\bm{G}))=1$ is maintained) in the BZBCs~Eq.\eqref{eq:BZBC&Vmat_defn}, thus breaking the periodic gauge choice, to make $\bm{u}_2$ continuous so that $\chi_{2,3}$ becomes well-defined} and will be quantized~\cite{Bouhon2020_NatPhys_ZrTe}. Doing so will however mean that the boundary integral in Eq.~\eqref{Eq:FullEuClass} does not vanish and thus a non-zero quantized Euler invariant no longer guarantees the quantization of $\bm{n}(\bm{k})$, i.e.~the contribution coming from the two-form curvature. This prevents extending, at least straight-forwardly, the linking protocol of Hopf fibers to non-orientable phases~\cite{Slager24NatCommNon-Abelian}.

Moreover, the non-trivial boundary conditions for Euler phases and their dynamics studied in our work can be experimentally probed with ultracold atoms. Indeed, kagome lattices have been already realized in optical lattices ~\cite{JoStamperKurn_PRL12}. The meronic Euler Hamiltonian of Eq.~\eqref{Eq:MeronH} that we examine is a kagome geometry with up to third next-nearest-neighbour tunnelings, where the longer-range couplings can be in principle achieved by using shallow optical potentials in experiments. Taking advantage of the greater tunability in these quantum simulators, different Euler phases can be explored by varying the ratios of hopping amplitudes along different directions and sublattice offsets. One needs to ensure the preservation of $\mathcal{C}_2$ symmetry, which is feasible given its rudimentary nature. 
The quench protocol can be then implemented by starting in the (trivial) atomic limit of kagome with deep lattice potentials and 
particles occupying only a single sub-lattice~\cite{Tarnowski19_NatCom}. The Hamiltonian can be then quenched by suddenly changing the laser parameters into a meronic Euler phase, while the evolving state can be measured via state tomography techniques that can be adopted to three-level systems~\cite{Even_Quench,Kemp_2022,Tarnowski19_NatCom} to extract the inverse images. Apart from optical lattices, investigating the quench dynamics of Type 2 Euler phases in trapped-ion setups offers another powerful approach to study non-trivial BZBCs and obstruction effects as it has been utilized for Type 1 Euler phases~\cite{EvenQuenchExper}, where implementing the Hamiltonian directly in the parameter space can provide unique advantages.

In conclusion, we here highlight the natural basis choices for the Bloch formalism with non-Bravais lattices and associated non-trivial Brillouin zone boundary conditions (BZBCs). Utilizing these definitions, we formulate a comprehensive classification of the possible lattice configurations for three-band Euler systems in two dimensions. Through examining the obstructed atomic limits of each case, we determine the parity of the Euler class that manifests in each of these possibilities. Our key highlight is that meronic phases with an odd invariant occur only when the BZBCs are non-trivial and anisotropic. Additionally, expanding upon existing schemes employing Hopf maps to determine the topological invariants in quench dynamics which have only considered periodic BZ boundaries so far, we establish a concrete formalism that accounts for non-trivial phase matrix BZBCs. This in turn not only caters to dynamical signatures of meronic Euler phases, but also demonstrates the observable effects of BZBCs and obstruction, which are indeed more than mere gauge artifacts. Our work paves the way for the investigation of novel multi-gap topological phases and understanding the role of obstruction and non-trivial BZBCs in non-Bravais lattices for other topologies.

\begin{acknowledgments}
	{\it Acknowledgments---} O.A.A.~acknowledges funding from the Saudi Arabian Ministry of Education through the Custodian of The Two Holy Mosques Research and Development Track scholarship program, and is grateful for their continued support during his academic journey. A.B.~acknowledges financial support from the Swedish Research Council (Vetenskapsradet) (Grant No. 2021-04681. R.-J.S.~acknowledges funding from an EPSRC ERC underwrite grant EP/X025829/1, and a Royal Society exchange grant IES/R1/221060. F.N.\"U.~acknowledges support from the Simons Investigator Award [Grant No.~511029], Trinity College Cambridge and the Royal Society under a University Research Fellowship Award [Grant No.~URF/R1/241667]. 
\end{acknowledgments}

\bibliography{references}

\clearpage
\newpage

\appendix

\section{Bloch Formalism for Non-Bravais Lattices, $\mathcal{C}_2\mathcal{T}$ Symmetry, and  Periodictization Indicated 1D Topological Obstructions in the Atomic Flag Limit}\label{ap_blochframe}

In this section, we review the Bloch formalism for non-Bravais 2D lattices, formally establishing the relation between the atomic degrees of freedom and the Brillouin zone periodicity of the corresponding Bloch Hamiltonian matrix, particularly in the presence of $\mathcal{C}_2\mathcal{T}$ symmetry (two-fold rotation composed with time reversal). We clarify the nomenclature and provide our \textit{periodictization procedure} that allows one to assess 1D topological obstructions in the atomic flag limit, and discuss subsequent implications on Euler topology of the system using the most natural form of the Bloch Hamiltonian matrix through particular basis and gauge choices. 

\subsection{Crystalline Atomic Degrees of Freedom}

We work with the lattice of a two-dimensional symmorphic crystal associated to a Bravais lattice (BL) defined as the group of vector translations obtained as integer linear combinations of two primitive vectors, $\boldsymbol{a}_1$ and $\boldsymbol{a}_2$, i.e.
\begin{equation}
    \boldsymbol{T} = \{\boldsymbol{R} =  n_1 \boldsymbol{a}_1 + n_2 \boldsymbol{a}_2 : n_1,n_2\in\mathbb{Z} \}\,.
\end{equation}
In each of the five inequivalent BLs in 2D, the unit cell is defined as the minimum space spanned by the primitive vectors needed to cover the whole lattice without overlapping, i.e. 
\begin{equation}
    u.c. = \{ t_1 \boldsymbol{a}_1 + t_2\boldsymbol{a}_2 : (t_1,t_2)\in (0,1]^2 \}\,. 
\end{equation}
The symmetry of a 2D crystal is defined by a {Layer group} $\mathcal{G}$, combining a Bravais lattice with one of the 32 three-dimensional crystallographic point groups $G$ (composed of rotations, reflections (mirrors), inversion and roto-inversion symmetries, and the identity element written as $E$ for the trivial case). Of the total of 80 layer groups, 43 are symmorphic and 37 non-symmorphic. Each element $\{g\vert \boldsymbol{R}\}$ of a symmorphic layer group can be separated into a translation of the 2D Bravais lattice, $\boldsymbol{R}\in \boldsymbol{T}$, and a point symmetry, $g \in G$, leading to a (in general) semi-direct product structure
\begin{equation}
    \mathcal{G} = \boldsymbol{T} \wedge G\,,
\end{equation}
with $\boldsymbol{T}$ a normal subgroup of $\mathcal{G}$ (i.e. $\{g\vert \boldsymbol{0}\}  \{E\vert \boldsymbol{R}\} \{g^{-1}\vert \boldsymbol{0}\} = \{ E\vert D_g \cdot \boldsymbol{R} \} \in \boldsymbol{T}$ for all $g\in G$, following the algebra of space group elements \cite{BradCrack}, where $D_g$ is the $O(3)$ rotation matrix associated to the point group element $g\in G$). The origin of the BL will always be chosen at a site of maximal point symmetry, i.e. a site invariant under the whole of $G$.

\subsubsection{Inequivalent Sublattice Degrees of Freedom}

Beyond the Bravais lattice sites, 2D crystals are composed of {\it sublattice} degrees of freedom, namely, a decoration of the unit cell by sites that cannot be joined from one to another through Bravais lattice translations. We represent the sublattice sites through their position vectors 
\begin{equation}
    \{\boldsymbol{r}_{i} \}_{i=1,\dots,N_{\text{sl}}}\,,
\end{equation}
such that each sublattice site appears only one time in the unit cell. Every sublattice site located away from the origin corresponds to a fractional position vector, that is~$\boldsymbol{r}_{i} = n_{i,1} \boldsymbol{a}_1 + n_{i,2}  \boldsymbol{a}_2$ with $0< \vert n_{i,1}\vert <1$ and $n_{i,2}=0$, or $0< \vert n_{i,2}\vert <1$ and $n_{i,1}=0$, or else $0< \vert n_{i,1}\vert, \vert n_{i,2}\vert <1$. 

We say that two sites of the crystal are {\it inequivalent} if the difference in their position vectors does not correspond to a vector of the Bravais lattice. Inversely, two sites belong to the same {\it equivalence class} if they are related by a translation of the Bravais lattice. Defining an equivalent class of sites among all sites $\{\boldsymbol{R}+\boldsymbol{r}_{i}\}_{\boldsymbol{R}\in\boldsymbol{T}}^{i=1,\dots,N_{\text{sl}}}$ as the left coset of a sublattice site $\boldsymbol{r}_i$ by the normal subgroup of Bravais lattice translation $\boldsymbol{T}$, i.e.
\begin{equation}
    [\boldsymbol{r}_i] =  \boldsymbol{r}_i+\boldsymbol{T}  =\left\{  
        \boldsymbol{r}_{i}+\boldsymbol{R} : \boldsymbol{R} \in \boldsymbol{T}
    \right\}\,,
\end{equation} 
we say that two sites, $\boldsymbol{R}+\boldsymbol{r}_{i}$ and $\boldsymbol{R}'+\boldsymbol{r}_{j}$, are equivalent if they belong to the same class of sites, i.e.
\begin{equation} 
    [\boldsymbol{R}+\boldsymbol{r}_{i}] = 
    [\boldsymbol{R}'+\boldsymbol{r}_{j}] \,.
\end{equation}
that is, if
\begin{equation} 
\begin{aligned}
    \boldsymbol{r}_{i} - \boldsymbol{r}_{j} = \boldsymbol{R}'' \in \boldsymbol{T}\,,
\end{aligned}
\end{equation}
in which case, they are counted as the same sublattice site degree of freedom. The requirement of non-repetition of sublattice sites within the unit cell means that the whole crystal can be decomposed into exactly $N_{\text{sl}}$ equivalence classes, that is one for each sublattice vector $\boldsymbol{r}_i$, i.e.
\begin{equation}
    \big\{ 
        [\boldsymbol{r}_i]
    \big\}_{i=1}^{ N_{\text{sl}} } \,.
\end{equation}

\subsubsection{Point Group Orbits of Equivalence Classes of Sites}

While, by definition, any two sublattice sites cannot be the images of one another under a translation by Bravais lattice vector, they can still be related by a point symmetry of the point group. An important concept associated to the sublattice sites is the {\it site symmetry group} $G_{\boldsymbol{r}_i}$, namely, the group of point symmetries $g\in G$ that leaves an equivalence class of sites, say $[\boldsymbol{r}_i]$, invariant, i.e.
\begin{equation}
    G_{\boldsymbol{r}_i} = \left\{g\in G : \left.^{g}[\boldsymbol{r}_i]\right. = [D_g\cdot \boldsymbol{r}_i] = [\boldsymbol{r}_i]
    \right\} \,,
\end{equation}
where $D_g$ is the $O(3)$ rotation matrix associated to the point symmetry $g$. Clearly, $G_{\boldsymbol{r}_i}$ must be a subgroup of $G$ for each sublattice site. Equivalently, acting on a site 
with a point symmetry that does not belong to its site symmetry group maps to another equivalence class of sites, i.e.
\begin{equation}
    \left.^{g}[\boldsymbol{r}_i]\right. = [\boldsymbol{r}_j] \neq [\boldsymbol{r}_i] \,, \;\text{if}\; g\notin G_{\boldsymbol{r}_i} \,.
\end{equation}

Let us now consider the partition of $G$ into left cosets of $G_{\boldsymbol{r}_i}$, i.e.
\begin{equation}
\label{eq_partition}
    G = \bigcup\limits_{m=1}^{n_{\boldsymbol{r}_i}}  g_{m} G_{\boldsymbol{r}_i} \,,
\end{equation}
where we choose $g_1\in G_{\boldsymbol{r}_i}$ (for instance $g_1=E$), such that the index of $G_{\boldsymbol{r}_i}$ in $G$, $ n_{\boldsymbol{r}_i} = [G : G_{\boldsymbol{r}_i}] $, is (by Lagrange theorem)
\begin{equation}
\label{eq_index_sitesymgroup}
    n_{\boldsymbol{r}_i} = \vert G \vert / \vert G_{\boldsymbol{r}_i} \vert \,,
\end{equation}
where `$\vert - \vert$' is the order (number of elements) of the finite group `$-$'. By definition, we have $g_m \notin G_{\boldsymbol{r}_i}$ for all $m\neq 1$. Then, the orbit of the site equivalence class $[\boldsymbol{r}_i]$ by $G$ is
\begin{equation}
\begin{aligned}
    \mathcal{O}_{\boldsymbol{r}_i} &= \bigl\{ \left.^{g}[\boldsymbol{r}_i]\right. : g\in G \bigr\} \,,\\
    &= \bigl\{ \left.^{g_m}[\boldsymbol{r}_i]\right.  \bigl\}_{m=1}^{n_{\boldsymbol{r}_i}} \,,
\end{aligned}
\end{equation}
with the elements $\{g_m\}_{m=1}^{n_{\boldsymbol{r}_i}}$ obtained in the partition Eq.\;(\ref{eq_partition}). The site symmetry groups of the sublattice sites belonging to the same orbit are all conjugated, namely, there is a $g_{m^{(ij)}} (\neq E)$ entering in the partition Eq.\;(\ref{eq_partition}), such that the site symmetry group $G_{\boldsymbol{r}_j}$ of the class
\begin{equation}
    [\boldsymbol{r}_j] = 
    \left.^{g_{m^{(ij)}}}[\boldsymbol{r}_i]\right.
    \in \mathcal{O}_{\boldsymbol{r}_i}
\end{equation}
is obtained through a conjugation
\begin{equation}
    G_{\boldsymbol{r}_j} = g_{m^{(ij)}} G_{\boldsymbol{r}_i} \, g_{m^{(ij)}}^{-1} \,,
\end{equation}
where $g_{m^{(ij)}} \notin G_{\boldsymbol{r}_i},  G_{\boldsymbol{r}_j}$, in which case we write $G_{\boldsymbol{r}_i}\sim G_{\boldsymbol{r}_j}$.  


\subsubsection{Wyckoff Positions}

For a given layer group, we can now decompose all the points in space into orbits of equivalence classes of sublattice sites or, scanning through the unit cell, 
\begin{equation}
    \mathcal{O}_{\boldsymbol{r}} = \bigl\{
        \left.^{g}[\boldsymbol{r}]\right. : g\in G
    \bigr\}
    \,,\;\text{for\,all}\;\boldsymbol{r}\in u.c.\,.
\end{equation}
Of any two points of the unit cell that belong to the same orbit (more precisely, whose equivalence classes belong to the same orbit), we say that they belong to the same {\it Wyckoff position} (WP). Furthermore, of any two points of the unit cell ($\boldsymbol{r}_i$, $\boldsymbol{r}_j$)
that belong to {\it distinct orbits} (i.e.~$^{g}[\boldsymbol{r}_i ] \neq [\boldsymbol{r}_j ]$ for all $ g\neq E$ in $G$) with the same site symmetry group, we again say that they belong to the same Wyckoff position if there is a path connecting them ($[\boldsymbol{r}_j\leftarrow \boldsymbol{r}_i]$) that preserves the site symmetry group, i.e. if
$G_{[\boldsymbol{r}_j\leftarrow \boldsymbol{r}_i]}=G_{\boldsymbol{r}_i}=G_{\boldsymbol{r}_j}$.


There is only a finite number of WPs for each layer group, which are usually labeled by a letter $\rho\in \{a,b,c,\dots,\mathcal{L}_E\} $, where the last letter $\mathcal{L}_E$ labels the single WP associated with the trivial site symmetry group $E$ (see below). Therefore, every sublattice site $\boldsymbol{r}_i$ and its associated orbit of equivalence class of sites $\mathcal{O}_{\boldsymbol{r}_i}$, both characterized by a site symmetry group $G_{\boldsymbol{r}_i} $, correspond to one WP $\rho$. WP $\rho$ is thus characterized by the point subgroup
\begin{equation}
    G_{\rho} = G_{\boldsymbol{r}_i} < G \,,
\end{equation}
that is the site symmetry group of any sublattice site that generates WP $\rho$. The WPs are usually denoted by $`n_{\rho} \rho'$ ($\rho=a,b,c,\dots,\mathcal{L}_E$) with the number
\begin{equation}
    n_{\rho} = \vert G \vert / \vert G_{\rho}\vert =  n_{\boldsymbol{r}_i} \,,
\end{equation}
that is the index of $G_{\rho} = G_{\boldsymbol{r}_i}$ in $G$ (defined in Eq.\;(\ref{eq_index_sitesymgroup})), with $\boldsymbol{r}_i$ any sublattice site that generates WP $\rho$, that is the number of sublattice sites in the orbit of equivalent classes of sites.  

Following the International Table for Crystallography \cite{ITCA,ITCE}, the labeling of WPs $\rho \in \{a,b,c,\dots,\mathcal{L}_E\}$ is always ordered from the site of maximal symmetry to the site of minimal symmetry, i.e.
\begin{equation}
    G_{a} = G > G_{b} > G_c > \cdots > G_{\mathcal{L}_E} \equiv E\,,
\end{equation}
with $\mathcal{L}_E$ given by the last letter of the alphabet needed to label all WPs. While there can be several WPs with the same site symmetry group, or with conjugated site symmetry groups, the WP associated to the trivial group $G_{\mathcal{L}_E} \equiv E$ is unique. Furthermore, some WPs are associated with a continuum of points in the unit cell, in which case their position vectors are parameterized by one or two variables (in 2D). For instance, WP $\mathcal{L}_E$ is generated by the general points $\boldsymbol{r}=(x,y)$ of the unit cell, that is, the points lying away from any symmetry center. We note that the layer group with the maximum number of WPs is the symmorphic $LG37 = pmmm$ with 18 WPs, the last of which is $8r$ with $G_{8r}=E$, in agreement with 8 being the number of elements in the orbit under point group $D_{2h}$, and $\vert D_{2h}\vert=8$. (The associated space group is $SG47$ with 27 WPs, the last of which is labeled with the capital letter $A$, i.e. $27A$ with $G_{27A}=E$, \cite{EvarestovSmirnov}.) 

In the following, we index all the occupied sublattice sites of the system according to their WPs, i.e. we write
\begin{equation}
\label{eq_sl_labeling}
    \{\boldsymbol{r}_{i}\}_{i=1}^{N_{\text{sl}}} = \bigcup\limits_{\rho\in I }
    \{
    \boldsymbol{r}_{\rho_{1} } , 
    \boldsymbol{r}_{\rho_{2} }, \dots, 
    \boldsymbol{r}_{\rho_{ n_{\rho} } } \}
    \,,
\end{equation}
where $I$ is the subset of labels of WPs occupied in the crystalline system, such that each WP is counted only one time, i.e. 
\begin{equation}
    I \subset \{a,b,c,\dots, \mathcal{L}_E\} \,,
\end{equation}
and such that 
\begin{equation}
    N_{\text{sl}} = \sum\limits_{\rho\in I} n_{\rho} \,.
\end{equation}

\subsubsection{Example for $LG80$}

Let us take the layer group $LG80$ as an example. It is obtained through the combination of the 2D hexagonal Bravais lattice, spanned by the primitive vectors $\boldsymbol{a}_1 = a (3/2,\sqrt{3}/2)$ and $\boldsymbol{a}_2 = a (-3/2,\sqrt{3}/2)$, with the point group $D_{6h}$. $LG80$ has 12 WPs, from WP $1a$ with the site symmetry group $G_{1a}=D_{6h}$, to WP $24l$ with the trivial site symmetry group $G_{24l}=E$. 

WP $a$ is associated to the sublattice position vector $\boldsymbol{r}_{a}  = \boldsymbol{0}$, the origin of the unit cell, and has a maximal site symmetry group $G_{a} = G = D_{6h}$, implying an orbit of a single sublattice site, i.e. $n_{a}= \vert G \vert /\vert G_{a}\vert = \vert D_{6h}\vert /\vert D_{6h}\vert=1$. 

WP $b$ corresponds to the sites forming the honeycomb lattice. It has the site symmetry group $G_{b} = D_{3h}$, implying a number of inequivalent sublattice sites $n_{b} = \vert D_{6h}\vert /\vert D_{3h}\vert = 2 $, with the position vectors 
\begin{equation}
\left\{
    \begin{aligned}
        \boldsymbol{r}_{b_1} &= \boldsymbol{a}_1/3+2\boldsymbol{a}_2/3\,,\\ 
        \boldsymbol{r}_{b_2} &= 2\boldsymbol{a}_1/3+\boldsymbol{a}_2/3 \,.
    \end{aligned}\right.
\end{equation}

WP $c$ corresponds to the sites forming the kagome lattice, where each site has a symmetry group $D_{2h}$, leading to a number of inequivalent sublattice sites $n_{c} = \vert D_{6h}\vert /\vert D_{2h}\vert = 3 $ (these are the sublattice sites labeled $\{A,B,C\}$ in the rest of the work) and the position vectors 
\begin{equation}
\left\{
    \begin{aligned}
        \boldsymbol{r}_{c_1} &=\boldsymbol{r}_A= \boldsymbol{a}_1/2\,,\\ 
        \boldsymbol{r}_{c_2} &=\boldsymbol{r}_B = 2\boldsymbol{a}_2 \,,\\
        \boldsymbol{r}_{c_3} &=\boldsymbol{r}_C = -\boldsymbol{a}_1/2-\boldsymbol{a}_2/2\,.
    \end{aligned}\right.
\end{equation}
 
If our system has atomic orbitals on the three types of WPs, we have $N_{\text{sl}}=6$ and we label the sublattice position vectors, according to Eq.\;(\ref{eq_sl_labeling}), as 
\begin{equation}
    \{\boldsymbol{r}_1,\dots ,  \boldsymbol{r}_6
    \} = \{ \boldsymbol{r}_{a} , \boldsymbol{r}_{b_1} , \boldsymbol{r}_{b_2}, 
    \boldsymbol{r}_{c_1} , \boldsymbol{r}_{c_2},
    \boldsymbol{r}_{c_3} 
    \}\,.
\end{equation}

\subsubsection{Orbital Degrees of Freedom}
In this work, we only consider integer orbital degrees of freedom, $l\in \{s,p,d,\dots\}$. (While we do not explicitly consider half-integer spin degrees of freedom in this work, our framework can be easily generalized to include Fermionic degrees of freedom.) We also assume that all (integer) orbital degrees of freedom are represented by the real basis functions of the irreducible representations of $SO(3)$, namely, by the cubic harmonics $\left\{ s,p_x,p_y,p_z,d_{3z^2-r^2},d_{x^2-y^2},d_{yz},d_{zx},d_{xy},\dots \right\}$. Each orbital $l$ has a dimension $n_l=2l+1$ ($n_s=1$, $n_p=3$, $n_d=5$, etc.), such that the orbital degrees of freedom realized at a WP $\rho$ can be indexed by 
\begin{equation}
    \left( 
    \bigl( l_{j'} 
    \bigr)_{j'=1}^{ n_{l} } 
    \right)_{l \in J_{\rho} }
    \,,
\end{equation}
where $J_{\rho}$ is the set of integer orbitals populating WP $\rho$, i.e.
\begin{equation}
    J_{\rho} \subset \{s,p,d,\dots \} \,.
\end{equation}
Allowing for the presence of multiple copies, say a number $n_{\rho,l}$, of orbital $l$ at the same WP $\rho$, we then index the principal orbital degrees of freedom by $\{ l^{\beta} \}^{\beta=1,\dots, n_{\rho,l} }$ with $l\in J_{\rho}$, and list them component-wise as
\begin{equation}
    \left( 
    \left(\Bigl( l^{\beta}_{j'} 
    \Bigr)_{j'=1}^{ n_{ l } }\right)_{\beta=1}^{n_{\rho,l}}
    \right)_{l\in J_{\rho}} \,.    
\end{equation}

\subsubsection{Combined Atomic Degrees of Freedom}

Combining all sublattice and orbital indices, the atomic degrees of freedom in each unit cell (which we will represent through localized Wannier functions, see below) are indexed by
\begin{equation}
\label{eq_labeling}
    \left(
    \left(
    \left(
    \left(
    \Bigl(
        w_{\rho_{j} , l^{\beta}_{ j'}}
    \Bigr)_{ j'=1}^{  n_l  } 
    \right)_{\beta=1}^{n_{\rho,l}}
    \right)_{l\in J_{\rho}}
    \right)_{j=1}^{n_{\rho}}
    \right)_{\rho\in I}
    \,,
\end{equation}
where $I\in \{a,b,c,\dots, \mathcal{L}_E\}$ is the subset of occupied WPs, $J\in\{s,p,d,\dots\}$ is the subset of occupied orbitals, $n_{\rho,l}$ fixes the number of copies of orbital $l$ located at WP $\rho$, and $n_{\rho}$ and $n_l$ are the intrinsic dimensionalities of WPs, and of integer angular momenta, respectively, both defined above. 

\subsubsection{Wannier States Basis}

We denote by 
\begin{equation}\label{eq:wannierbasis}
    \Bigl\{\vert w_{\alpha_i} , \boldsymbol{R} + \boldsymbol{r}_{\alpha_i}\rangle \Bigr\}_{i=1}^{M}
\end{equation}
the $M$ atomic-like Wannier states associated to the Wannier functions $\langle \boldsymbol{x}\vert   w_{\alpha_i} , \boldsymbol{R} + \boldsymbol{r}_{\alpha_i} \rangle = w_{\alpha_i} (\boldsymbol{x}- \boldsymbol{R} - \boldsymbol{r}_{\alpha_i})$ localized at the position vectors
\begin{equation}
    \Bigl\{\boldsymbol{R} + \boldsymbol{r}_{\alpha_i }
    \Bigr\}_{i=1}^M    \,,
\end{equation}
within the unit cell centered at $\boldsymbol{R}\in\boldsymbol{T}$, and representing the $M$ atomic degrees of freedom that combine sublattice degrees of freedom and the orbital degrees of freedom. Clearly, the counting of degrees of freedom must match with
\begin{equation}
    N_{\text{sl}} = \sum\limits_{\rho\in I} n_{\rho} \,,
\end{equation}
and
\begin{equation}
    M = \sum\limits_{\rho\in I} \sum\limits_{l\in J} n_{\rho} \, n_{\rho,l} \, n_l \,.
\end{equation}

On top of separating the lattice degrees of freedom from the rest, Eq.\;(\ref{eq_labeling}) imposes a specific ordering of the Wannier states basis. Indeed, by setting
\begin{subequations}
    \label{eq_indices}
\begin{equation}
\begin{aligned}
    \vert \boldsymbol{w},\boldsymbol{R}\rangle &= 
    \left(
        \vert w_{\alpha_1},\boldsymbol{R}\rangle \cdots 
        \vert w_{\alpha_{M}} ,\boldsymbol{R}\rangle
    \right) \\
    &= \left(
        \vert \boldsymbol{w}^{(1)},\boldsymbol{R}\rangle \cdots 
        \vert \boldsymbol{w}^{(N_{\text{sl}})},\boldsymbol{R}\rangle
    \right) \\
    &= \Bigl(
        \vert \boldsymbol{w}_{\rho_1},\boldsymbol{R}\rangle \cdots 
        \vert \boldsymbol{w}_{\rho_{n_{\rho}}},\boldsymbol{R}\rangle
    \Bigr)_{\rho\in I} \\
    = \Bigl(
        \vert \boldsymbol{w}_{a_1},\boldsymbol{R}\rangle & 
        \cdots 
        \vert \boldsymbol{w}_{a_{n_{a}}},\boldsymbol{R}\rangle ~
        \vert \boldsymbol{w}_{b_1},\boldsymbol{R}\rangle \cdots 
        \vert \boldsymbol{w}_{b_{n_{b}}},\boldsymbol{R}\rangle ~
        \cdots
    \Bigr)
    \,,
\end{aligned}
\end{equation}
with, for each $\rho\in I$ and $j\in \{ 1,\dots, n_{\rho}\}$,
\begin{equation}
\begin{aligned}
    & \vert \boldsymbol{w}_{\rho_j},\boldsymbol{R}\rangle =  \\
    & \biggl(
        \left\vert w_{(\rho_j , l^1_1 )  } , \boldsymbol{R}+\boldsymbol{r}_{ \rho_j } \right\rangle  
          \cdots
        \left\vert w_{(\rho_j , l^1_{n_l} )} , \boldsymbol{R}+\boldsymbol{r}_{\rho_j} \right\rangle  \\
    & \quad \left\vert w_{(\rho_j , l^2_1 )  } , \boldsymbol{R}+\boldsymbol{r}_{ \rho_j } \right\rangle  
          \cdots 
        \left\vert w_{(\rho_j , l^2_{n_l} )} , \boldsymbol{R}+\boldsymbol{r}_{\rho_j} \right\rangle \\
    & \quad\quad\quad\quad \vdots\\ 
      & \quad\quad \left\vert w_{(\rho_j , l^{n_{\rho,l}}_1 )  } , \boldsymbol{R}+\boldsymbol{r}_{ \rho_j } \right\rangle  
          \cdots 
        \left\vert w_{(\rho_j , l^{n_{\rho,l}}_{n_l} )} , \boldsymbol{R}+\boldsymbol{r}_{\rho_j} \right\rangle
    \biggr)\,,
\end{aligned}
\end{equation}
\end{subequations}
we have a representation of the total Hilbert space 
\begin{equation}
    \mathcal{H} = 
    \mathcal{H}_{\text{BL}} \otimes 
    \mathcal{H}_{\text{sl}} \otimes
    \mathcal{H}_{\text{orb}}\,.
\end{equation}
We can then build the corresponding Hamiltonian operator in the Wannier states basis
\begin{equation}
    \mathcal{H} = \sum\limits_{\boldsymbol{R},\boldsymbol{R}'} \vert \boldsymbol{w}, \boldsymbol{R} \rangle \cdot H_{W}(\boldsymbol{R}'-\boldsymbol{R}) \cdot 
    \langle \boldsymbol{w}, \boldsymbol{R}' \vert 
\end{equation}
with the Wannier Hamiltonian matrix
\begin{equation}
\left[ H_{W}
    (\boldsymbol{R}'-\boldsymbol{R})\right]_{\alpha_i\alpha_j} = 
    \left[ H_{W}
    (\boldsymbol{R}'-\boldsymbol{R})\right]_{(\rho_j,l^{\beta}_{j'}) (\tilde{\rho}_{\tilde{j}},\tilde{l}^{\tilde{\beta}}_{\tilde{j}'})} ,
\end{equation}
where the correspondence between indices is given through Eqs.\;(\ref{eq_indices}).

\subsection{Zak Bloch-basis, Zak phase, and First Periodictization}

We now define a Bloch basis that represents the inequivalent atomic degrees of freedom within each unit cell, while factoring the Bravais lattice degrees of freedom through a discrete Fourier transform. We call the following basis the {\it Zak basis} for reasons that will become clear below,
\begin{equation}
\label{eq_Bloch_basis_Zak}
\left\{
    \begin{aligned}
    \vert \varphi^Z_{\alpha_i},\boldsymbol{k} \rangle &= \dfrac{1}{\sqrt{N_{\alpha_i}}}\sum\limits_{\boldsymbol{R}_m \in \text{BL}} e^{\text{i} \boldsymbol{k}\cdot\left(\boldsymbol{R}_m +\boldsymbol{r}_{\alpha_i}\right)} 
    \vert w_{\alpha_i},\boldsymbol{R}_m+\boldsymbol{r}_{\alpha_i} \rangle \,, \\
    \vert \boldsymbol{\varphi}^Z,\boldsymbol{k}\rangle &= \left(\vert \varphi^Z_{\alpha_1},\boldsymbol{k}\rangle ~\cdots~ \vert \varphi^Z_{\alpha_M},\boldsymbol{k}\rangle \right) \,,
    \end{aligned}\right.
\end{equation}
with $N_{\alpha_i}$ the number of $\alpha_i$ states in the whole system (effectively, we take a finite lattice with periodic boundary conditions and then send the number of discrete sites to infinity) and $\boldsymbol{k}$ the quasimomentum designing a point of the BZ. It is important to notice the inclusion of the phase factor $e^{\text{i} \boldsymbol{k}\cdot\boldsymbol{r}_{\alpha} }$ in the Fourier transform. This is a natural choice because it leads to an almost canonical definition of a Bloch connection from the frame of Bloch basis states, as
\begin{equation}\label{eq:Bloch_1form}
\begin{aligned}
    \mathcal{A}^Z(\boldsymbol{k}) &= \langle \boldsymbol{\varphi}^Z , \boldsymbol{k} \vert \partial_{k^{\mu}} \vert \boldsymbol{\varphi}^Z , \boldsymbol{k} \rangle dk^{\mu}\,.
\end{aligned}
\end{equation}
We then find that a change of the Bravais lattice origin $ (0,0) \mapsto \boldsymbol{\Delta}$, i.e. $\boldsymbol{R}_m = \boldsymbol{R}_m' +  \boldsymbol{\Delta}$, leads to a change of the Bloch connection by
\begin{equation}
    \mathcal{A'}^{Z}(\boldsymbol{k}) = \mathcal{A}^{Z}(\boldsymbol{k}) - \text{i} \boldsymbol{\Delta}\cdot d\boldsymbol{k}\,,
\end{equation}
namely, the difference is a scalar-valued closed 1-form. We thus readily obtain that the Bloch curvature 2-form in the Zak's gauge, $\mathcal{F}^Z = d\mathcal{A}$, is invariant under a change of Bravais lattice origin (since $d^2=0$), see \cite{Fruchart_2014} for a more detailed discussion. Furthermore, this choice of Bloch basis leads to a direct equivalence between the Zak phases (i.e.~the Berry phase computed over the non-contractible paths of the Brillouin zone torus, see below) and the expectation values of the position operator, the so called ``band centers'' or ``Wannier centers'', see Section \ref{sec_Zakphase} and the references therein. 

We now define the Bloch Hamiltonian operator
\begin{equation}
\left\{
    \begin{aligned}
    \mathcal{H} &= \sum\limits_{\boldsymbol{k}} \mathcal{H}_{\boldsymbol{k}} \,, \\
    \mathcal{H}_{\boldsymbol{k}} &= \vert \boldsymbol{\varphi}^Z,\boldsymbol{k}\rangle   \cdot
    H^Z(\boldsymbol{k}) \cdot 
    \langle \boldsymbol{\varphi}^Z ,\boldsymbol{k}\vert \,.
    \end{aligned}
\right.
\end{equation}
where the elements of the Bloch Hamiltonian matrix $H^Z(\boldsymbol{k})$ are given as two-dimensional (finite, see the discussion below) Fourier series in the $(k_1,k_2)$-components of the quasimomentum, i.e.
\begin{equation}
\begin{aligned}
    \left[H^Z(\boldsymbol{k})\right]_{ij} &= \langle \varphi^Z_{\alpha_i}, \boldsymbol{k} \vert \hat{h} \vert \varphi_{\alpha_j}^Z,\boldsymbol{k}\rangle \,,\\
    &= \sum\limits_{\boldsymbol{R}\in \text{BL}} e^{\text{i} \boldsymbol{k}\cdot (\boldsymbol{R}+\boldsymbol{r}_{\alpha_j}-\boldsymbol{0}-\boldsymbol{r}_{\alpha_i})} 
    t_{\alpha_i\alpha_j}(\boldsymbol{0}-\boldsymbol{R})\,,
\end{aligned}
\end{equation}
with $\hat{h}$ the one-body operator associated to the static Schr{\"o}dinger equation with periodic potential (e.g.~given by the Kohn-Sham Hamiltonian from ab initio). The coefficients $t_{\alpha_i\alpha_j}(\boldsymbol{0}-\boldsymbol{R})$ give the tunneling amplitudes between the $\alpha_i$ and $\alpha_j$ atomic Wannier states located at unit cells separated by the distance $\vert \boldsymbol{R}\vert$. The assumption that the Fourier series is finite, i.e.~$\boldsymbol{R}$ in the sum only runs through a finite number of cells around the origin $\boldsymbol{0}$, which is motivated physically by the fact that atomic tunneling amplitudes die exponentially fast for increasing hopping distances, implies that each matrix element $[H^Z(\boldsymbol{k})]_{ij}$ is an analytical function of $k_1$ and $k_2$. The nontrivial topology, and the associated singularities of the Bloch eigenstates, come from the spectral separation between groups of bands. See the Appendix of Ref.~\cite{Chen2023-ui} for a more detailed presentation of the formalism of the modeling.  

The band structure is then obtained through the solution of the Bloch eigenvalue problem 
\begin{equation}
    H^Z(\boldsymbol{k}) u^Z_n(\boldsymbol{k}) =  u^Z_n(\boldsymbol{k}) \, E_n(\boldsymbol{k})  
    \,,
\end{equation}
with the $n$-th energy eigenvalue $E_n(\boldsymbol{k})$ ($n=1,\dots,M$) and the associated $n$-th Bloch eigenvector $u^Z_n(\boldsymbol{k}) \in \mathbb{C}^M$. In the following we will always assume that the energy eigenvalues are labeled in  order of increasing energy, i.e.
\begin{equation}
    E_{n}(\boldsymbol{k}) \leq E_{n+1}(\boldsymbol{k})\,,
\end{equation}
for all $n\in \{1,\dots,M-1\}$.

\subsubsection{Periodicity of the Bloch Hamiltonian Matrix in Reciprocal Space}\label{sec_periodicity_Blochop}

From the invariance of the Bloch Hamiltonian operator under translation by reciprocal lattice vectors, i.e. 
\begin{equation}
    \mathcal{H}_{\boldsymbol{k}+\boldsymbol{G}} = \mathcal{H}_{\boldsymbol{k}}\,,    
\end{equation}
with $\bm{G} = m_1 \bm{b_1} + m_2 \bm{b_2}$ ($m_1,m_2 \in \mathbb{Z}$), where $\{\bm{b}_{1},\bm{b}_{2}\}$ are the two reciprocal primitive lattice vectors, we have
\begin{equation}\label{Eq:Vmat_Origin}
    V^Z(\boldsymbol{G}) \cdot H^Z(\boldsymbol{k}+\boldsymbol{G}) \cdot V^Z(\boldsymbol{G})^{\dagger} = H^Z(\boldsymbol{k}) \,,
\end{equation}
with the \textit{atomic phase matrix} (that we also call the shift matrix) as also defined in the main text:
\begin{equation}\label{Eq:Vmat_defn_appendix}
    V^Z(\boldsymbol{G}) = \text{diag}\left(
         e^{\text{i} \boldsymbol{G}\cdot \boldsymbol{r}_{\alpha_1} },
         \cdots
         , e^{\text{i} \boldsymbol{G}\cdot \boldsymbol{r}_{\alpha_M} } 
         \right)   \,,
\end{equation}
such that 
\begin{equation}
    \vert \boldsymbol{\varphi}^Z, \boldsymbol{k}+\boldsymbol{G}\rangle = \vert \boldsymbol{\varphi}^Z, \boldsymbol{k}\rangle \cdot V^Z(\boldsymbol{G})\,.
\end{equation}

We will call this relation the {\it Brillouin zone boundary condition} of the Bloch Hamiltonian in this work. Since $V^Z(\boldsymbol{G})$ is not an identity matrix in general as $\boldsymbol{r}_{\alpha_i}$ can be non-zero, we see that the Bloch Hamiltonian matrix $H^Z(\boldsymbol{k})$ is not necessarily periodic in reciprocal space. It also implies for the Bloch eigenvectors
\begin{equation}
\label{eq_shifted_eigenvec}
    u^Z_n(\boldsymbol{k}+\boldsymbol{G}) = s_n 
    V^Z(\boldsymbol{G})^{\dagger}\cdot u^Z_n(\boldsymbol{k})
    \, , \quad s_n=\pm1 \,,
\end{equation}
where $s_n$ represents the gauge sign freedom for real eigenvectors. This fixes the gauge signs of the Bloch eigenvectors at the shifted quasimomentum $\boldsymbol{k}+\boldsymbol{G}$ by lattice translations $\boldsymbol{G}$ from the gauges chosen at $\boldsymbol{k}$. 
The choice of gauge sign $s_n=1$ for all $n$ is called the {\it periodic gauge} and is used later on in the definition of the Zak phases. We emphasize that the existence of periodic Bloch eigenvectors along the direction of $\boldsymbol{G}$ is also determined by the shift matrix $V^Z(\boldsymbol{G})$, according to Eq.\;(\ref{eq_shifted_eigenvec}). 

This procedure of choosing the canonical Zak-basis is the first step of \textit{periodictization}. The Bloch eigenvectors in the atomic flag limit are periodic up to the diagonal atomic phase matrix $V^Z(\boldsymbol{G})$ which we just saw represents the shift by a reciprocal vector $\bm{G}$. We further discuss in Section~\ref{sec_maxper} below the conditions for the existence of a subset of parallel-transported real eigenvectors that are also periodic in the atomic flag limit,howcasing that the full set of real eigenvectors is maximally, but not precisely fully, periodic to maintain the 1D topology encoded in the Zak phases and keep the physical interpretation of correspondence with Wannier centers as much as possible.

\subsubsection{Zak Phase in the Zak Basis}\label{sec_Zakphase}

We now define the Zak phase for a group of $p_{\nu}$ bands ($p_{\nu}=1,2,\dots$) associated to the $p_{\nu}$ Bloch eigenvectors $\{u_{{\nu}+1}(\boldsymbol{k}),\dots, u_{{\nu}+p_{\nu}}(\boldsymbol{k})\}$ (labeling the successive groups of bands in the order of increasing energy by $\nu=1,2,\dots$) over one commensurate direction of the reciprocal space, say from a quasimomentum $\boldsymbol{k}_0$ to the shifted point $\boldsymbol{k}_0+\boldsymbol{G}$, where $\boldsymbol{G}$ is the minimal reciprocal lattice vector in that direction. The $\nu$-th group of bands is well defined when it satisfies the adiabatic conditions $E_{\nu}(\boldsymbol{k}) < E_{\nu+1}(\boldsymbol{k})$ and $E_{\nu+p_{\nu}}(\boldsymbol{k}) < E_{\nu+p_{\nu}+1}(\boldsymbol{k})$ for all $\boldsymbol{k}\in [\boldsymbol{k}_0,\boldsymbol{k}_0+\boldsymbol{G}]$. While we write the definition directly in terms of the Bloch eigenvectors, it can be shown that it is fully compatible with the definition of the Zak phase from the cell-periodic part of the Bloch eigenfunctions, given as (by the Boch theorem)
\begin{equation}
    u_{n,\boldsymbol{k}}(\boldsymbol{r})
    = e^{-\text{i}\,\boldsymbol{k}\cdot \boldsymbol{r}} \, \psi_{n,\boldsymbol{k}}(\boldsymbol{r}) \,,
\end{equation}
$n=1,\dots,M$, with the Bloch eigenfunctions defined as the position-representation, $\psi_{n,\boldsymbol{k}}(\boldsymbol{r}) = \langle \boldsymbol{r}\vert \psi_{n} ,\boldsymbol{k}\rangle $, of the Bloch eigen{\it states}
\begin{equation}
\vert \psi_{n} ,\boldsymbol{k}\rangle = \vert \boldsymbol{\varphi}^Z, \boldsymbol{k}\rangle \cdot u^Z_n(\boldsymbol{k}) \,.
\end{equation}
The Bloch eigenstates are themselves the solutions of the eigenvalue problem written at the level of the Bloch Hamiltonian {\it operator}, i.e.
\begin{equation}
    \mathcal{H}_{\boldsymbol{k}} \vert \psi_{n} ,\boldsymbol{k}\rangle = \vert \psi_{n} ,\boldsymbol{k}\rangle \,E_n(\boldsymbol{k})    \,.
\end{equation}

Writing the  partial frame of $p_\nu$ Bloch eigenvectors
\begin{equation}
    \mathcal{U}^Z_{\nu}(\boldsymbol{k}) = 
    \left(u^{Z}_{\nu+1}(\boldsymbol{k})\cdots
    u^{Z}_{\nu+p_{\nu}}(\boldsymbol{k})
    \right)\,,
\end{equation}
and the Wilson loop operator of the given band-subspace over the path $[\boldsymbol{k}_0,\boldsymbol{k}_0+\boldsymbol{G}]$, i.e.
\begin{equation}
W^Z_{\nu,[\boldsymbol{k}_0+\boldsymbol{G}\leftarrow \boldsymbol{k}_0]} = \prod\limits_{\boldsymbol{k}}^{[\boldsymbol{k}_0+\boldsymbol{G}\leftarrow \boldsymbol{k}_0]} \mathcal{U}^Z_{\nu}(\boldsymbol{k})\cdot \mathcal{U}^Z_{\nu}(\boldsymbol{k})^{\dagger}\,,
\end{equation}
we define the Zak phase by
\begin{subequations}
\label{eq_Zak_phase_definition}
\begin{equation}
    e^{-\text{i} \gamma_{\nu}[\boldsymbol{k}_0+\boldsymbol{G}\leftarrow \boldsymbol{k}_0]} = \det \mathcal{W}_{\nu, [\boldsymbol{k}_0+\boldsymbol{G}\leftarrow \boldsymbol{k}_0]} \,,
\end{equation}
where the Wilson loop matrix $\mathcal{W}_{\nu, [\boldsymbol{k}_0+\boldsymbol{G}\leftarrow \boldsymbol{k}_0]} \in \mathsf{U}(p_{\nu})$ is obtained from the Wilson loop operator through 
\begin{equation}
\begin{aligned}
\mathcal{W}_{\nu, [\boldsymbol{k}_0+\boldsymbol{G}\leftarrow \boldsymbol{k}_0]} &= 
    \mathcal{U}^Z_{\nu}(\boldsymbol{k}_0+\boldsymbol{G} )^{\dagger}
     \cdot W^Z_{\nu,[\boldsymbol{k}_0+\boldsymbol{G}\leftarrow \boldsymbol{k}_0]}
     \cdot \mathcal{U}^Z_{\nu}(\boldsymbol{k}_0) \,,\\
     = \mathcal{U}^Z_{\nu}(\boldsymbol{k}_0)^{\dagger} & \cdot 
     V^Z(\boldsymbol{G})
     \cdot W^Z_{\nu,[\boldsymbol{k}_0+\boldsymbol{G}\leftarrow \boldsymbol{k}_0]}
     \cdot \mathcal{U}^Z_{\nu}(\boldsymbol{k}_0) ,
\end{aligned}
\end{equation}
\end{subequations}
where we have used the periodic gauge as by fixing $s_n=1$ in Eq.\;(\ref{eq_shifted_eigenvec}) for $n\in \nu+1, \dots, \nu+p_{\nu}$. 

As already mentioned above, the Zak phase obtained in the Zak basis is equivalent to the expectation value in the band-Wannier basis of the position operator (the component of the position vector projected on the direction of the commensurate quasimomentum path), as first pointed out by Zak \cite{Zakbandcenter,Zakphase} who called it band center or, as in \cite{Van2,Wi1}, Wannier center. We show below that the Zak phases directly capture the position of atomic orbitals in the atomic flag limit.

\subsubsection{Atomic Flag Limit} \label{secAppx:AtomicFlagLim}

We define the {\it atomic flag limit} as an atomic limit, by setting the Bloch Hamiltonian matrix to a constant diagonal matrix, for which the onsite energies of the atomic orbitals are all given a different value, i.e. we set
\begin{equation}
\left\{
\begin{aligned}
    H^Z &= \text{diag}(\epsilon_{\alpha_1}, \dots,  \epsilon_{\alpha_M}) \,,\\
    & \epsilon_{\alpha_1} < \dots <  \epsilon_{\alpha_M} \,.
\end{aligned}\right.
\end{equation}
We call it ``flag'' because each band is separated from the others by an energy gap (this leads to a space of Hamiltonian with the geometry of a flag manifold, contrary to the situation of a single energy gap for which the space of Hamiltonian is a Grassmannian). The Bloch eigenvalues are thus
\begin{equation}
    E_n(\boldsymbol{k}) = \epsilon_{\alpha_n} \,,
\end{equation}
with the Bloch eigenvectors given by the constant coordinate vectors 
\begin{equation}
    [u_n(\boldsymbol{k})]_j = \delta_{n,j} \,,\;\text{for}\;n,j=1,\dots,M \,.
\end{equation}
It follows from the above definition that the Zak phase of each atomic band is readily given by it by the sublattice site position of the corresponding atomic site, i.e. taking Eq.\;(\ref{eq_Zak_phase_definition} with $\boldsymbol{k}_0=\boldsymbol{0}$ we get 
\begin{equation}
\label{eq_Zak_flag_atomic_limit}
    e^{\text{i} \gamma_{n}[\boldsymbol{G}] } = 
    [V^Z(\boldsymbol{G})]_n = 
    e^{\text{i} \boldsymbol{G}\cdot \boldsymbol{r}_{\alpha_n} }
    \,,
\end{equation}
where $\boldsymbol{r}_{\alpha_n}$ is the sublattice location of the atomic orbital $\alpha_n$.

\subsection{$\mathcal{C}_2\mathcal{T}$ Symmetry, Reality Condition and Takagi Factorization}\label{appsec:A_c2t_and_reality}


In general, the Bloch Hamiltonian matrix $H^Z(\boldsymbol{k})$ is complex and there is no one-dimensional topology associated with its lack of periodicity under translations by reciprocal lattice vectors (we discard here topological features linked to one-dimensional non-symmorphic symmetries, such as multi-band pairing as discussed e.g. in \cite{Chen2023-ui}). The situation is different when the hermitian Bloch Hamiltonian matrix can be brought into a real symmetric matrix through a change of Bloch basis. This happens when the system satisfies the so-called {\it reality condition}, that is, when there exists an anti-unitary symmetry $A$ (that is, for us, a symmetry combining a unitary operator with complex conjugation, such as time reversal and particle-hole symmetries in quantum mechanics) that squares to identity ($A^2=\mathbb{1}$) and leaves the quasimomentum unchanged, i.e.~$\boldsymbol{k}\mapsto -D_A\cdot \boldsymbol{k} = \boldsymbol{k}$, where $D_A$ is the rotation matrix associated with the action of the unitary part of the symmetry $A$. 

A common realization of the reality condition in crystaline systems is given by the symmetry $\mathcal{C}_2\mathcal{T}$ which combines $\mathcal{C}_{2,z}$ rotation ($\pi$-rotation around the vertical $\hat{z}$-axis perpendicular to the basal plane of the 2D system, crossing the origin of the unit cell) and time reversal symmetry $\mathcal{T}$. Remarkably, this is independent of whether the degrees of freedom are bosonic or fermionic, since $[\mathcal{C}_2\mathcal{T}]^2=\mathbb{1}$ for both cases, as can be easily verified \footnote{Note that the reality condition can be obtained from $\mathcal{P}\mathcal{T}^+$ combining inversion $\mathcal{P}$ and bosonic time reversal symmetry ($[\mathcal{T}^+]^2=\mathbb{1}$).}. 


As we show below (Sec.\;\ref{sec_reality_cond}), any system satisfying the reality condition can be transformed into a Bloch basis ($\boldsymbol{\varphi}^Z\mapsto \widetilde{\boldsymbol{\varphi}}$) that makes the Bloch Hamiltonian matrix real, i.e. $H^Z(\boldsymbol{k})\mapsto \widetilde{H}(\boldsymbol{k}) \in \mathbb{R}^M\times \mathbb{R}^M$ \cite{Bouhon2020_NatPhys_ZrTe}. In that case, the Bloch eigenvectors can thus be mapped to real vectors, such that any group of bands separated from the other bands by an energy gap from above and from below and taken over a commensurate 1D direction of the reciprocal space (i.e., connecting two distinct sites of the reciprocal Bravais lattice) defines a {\it real} vector bundle. Real vector bundles over a one-dimensional base space (here the quasimomentum path crossing the BZ) can be topologically characterized according to their orientability. The stable $\mathbb{Z}_2$ topological invariant associated to the orientability of real vector bundles is the {\it first} Stiefel-Whitney class \cite{BBS_nodal_lines} or, equivalently, the $\mathbb{Z}_2$-quantized Zak phase \cite{Zakphase}. We rederive below the $\mathbb{Z}_2$-quantization of the Zak phase for systems satisfying the reality condition.

\subsubsection{$\mathcal{C}_2\mathcal{T}$ Action on The Sublattice Basis}\label{sec_sublattices_C2}

Crucially, only $\mathcal{C}_2$ in $\mathcal{C}_2\mathcal{T}$ acts on the position operator. This implies that the underlying crystal structure must be effectively $\mathcal{C}_2$-symmetric, even though the Hamiltonian itself is not necessarily symmetric under $\mathcal{C}_2$. As a consequence, we can distinguish the WPs that are invariant under $\mathcal{C}_{2}$ symmetry (more precisely, those for which the corresponding equivalence classes, i.e.~up to a Bravais lattice vector, are invariant), which we call the $\mathcal{C}_{2}$ centers and write them with a star $\rho^*$, while those that are not are written without a star. We also note that here we only consider the $\mathcal{C}_2^{\perp}$ symmetry that is perpendicular to the basal plane of the 2D system, since in-plane $\mathcal{C}^{\parallel}_2$ symmetries are only associated to one-dimensional regions of the BZ satisfying the reality condition. In the chosen axis, we have here $\mathcal{C}_{2}^{\perp}=\mathcal{C}_{2}^{z}$ which we simply keep writing as $\mathcal{C}_2$.

By definition, the invariance group of a WP $\rho^*$ that is a $\mathcal{C}_2$ center contains $\mathcal{C}_2$ as a subgroup, i.e.
\begin{equation}
    \mathcal{C}_2 < G_{\rho^*}\,. 
\end{equation}
The action of $\mathcal{C}_2\mathcal{T}$ on the equivalence class of sublattice sites $[\boldsymbol{r}_{\rho^*_j}]$ belonging to $\mathcal{C}_2$-symmetric WP $\rho^*$ is thus simply 
\begin{equation}
    \left.^{\mathcal{C}_2\mathcal{T}}[\boldsymbol{r}_{\rho^*_j}]\right. = [D_2\cdot \boldsymbol{r}_{\rho^*_j}] = [\boldsymbol{r}_{\rho^*_j}] \,,
\end{equation}
given that $\mathcal{T}$ acts trivially on position operators. That is, the representation of $\mathcal{C}_2\mathcal{T}$ in the sublattice space is trivial. In general, when WP $\rho$ is not a $\mathcal{C}_2$ center, $n_{\rho}$ must be even, and we get
\begin{equation}
\begin{aligned}
    \left.^{\mathcal{C}_2\mathcal{T}}\left(
        [\boldsymbol{r}_{\rho_1}]   \cdots
         [\boldsymbol{r}_{\rho_{n_{\rho}}}] 
    \right)\right. &= \left(
        [D_2\cdot \boldsymbol{r}_{\rho_1}]   \cdots
         [D_2\cdot\boldsymbol{r}_{\rho_{ n_{\rho} } 
 }] 
    \right) \\
    & = \left(
        [\boldsymbol{r}_{\rho_1}]   \cdots
         [\boldsymbol{r}_{\rho_{n_{\rho}}}] 
    \right) \cdot \sigma_{\rho}(\mathcal{C}_2)\,,
\end{aligned}
\end{equation}
where $ \sigma_{\rho}(\mathcal{C}_2)$ is a twofold (involutary) permutation matrix, i.e. $\sigma_{\rho}(\mathcal{C}_2)^2 = \mathbb{1}$. 

We now further order the Bloch basis Eq.\;(\ref{eq_Bloch_basis_Zak}) beyond the WP-sublattice-orbital order given to the Wannier basis in Eq.\;(\ref{eq_indices}). For any WP $n_{\rho}\rho$ ($\rho\in\{a,b,c,\dots\}$) that is not a $\mathcal{C}_2$-center ($n_{\rho}$ is even), we decompose all the associated sublattice positions, $\{\boldsymbol{r}_{\rho_1},\dots,\boldsymbol{r}_{\rho_{n_{\rho}}}\}$, in pairs that are images of one-another under $\mathcal{C}_2$, i.e.~we choose the labeling such that 
\begin{equation}
    ^{\mathcal{C}_2}
    [ \boldsymbol{r}_{2j-1} ] = [D_2\cdot \boldsymbol{r}_{2j-1}] = [\boldsymbol{r}_{2j}] \,,
\end{equation}
for $j \in \{1,\dots,n_{\rho}/2\}$. The representation of $\mathcal{C}_2 \mathcal{T}$ in the basis of equivalent classes of positions associated to a non-$\mathcal{C}_2$-symmetric center thus has the following structure 
\begin{equation}
\left\{
\begin{aligned}
    ^{\mathcal{C}_2\mathcal{T}}
    \left([ \boldsymbol{r}_{1} ]\cdots [ \boldsymbol{r}_{n_{\rho}} ]\right) &=  \left([ \boldsymbol{r}_{1} ]\cdots [ \boldsymbol{r}_{n_{\rho}} ]\right) \cdot U^{(\rho)}_{\text{sl},2'} \,,\\
    U^{(\rho)}_{\text{sl},2'} &= 
    \mathbb{1}_{n_{\rho}/2\times n_{\rho}/2} \otimes \sigma_x\,.
\end{aligned}\right.
\end{equation}
On the other hand, the $\mathcal{C}_2\mathcal{T}$ representation in the basis of $\mathcal{C}_2$-centers $\rho^*$ is simply the identity matrix. 

We emphasize that the geometry of the Bravais lattice chosen to tile the $\mathcal{C}_2$-symmetric lattice of atomic orbitals impacts which sites lie at a $\mathcal{C}_2$ center and which do not. In the following, we assume that the BL is defined such that it faithfully captures the underlying $\mathcal{C}_2$ symmetry, with the maximal number of atomic orbitals being at a $\mathcal{C}_2$ center site.

\subsubsection{$\mathcal{C}_2\mathcal{T}$-Representation and Reality Condition}\label{sec_reality_cond}


$\mathcal{C}_2\mathcal{T}$ symmetry (also written as $2'$ in the International Tables for Crystallography) satisfies the reality condition, since $[\mathcal{C}_2\mathcal{T}]^2=\mathbb{1}$ and both $\mathcal{C}_2$ and $\mathcal{T}$ take $\boldsymbol{k}$ to $-\boldsymbol{k}$ in the basal 2D Brillouin zone, such that $\mathcal{C}_2\mathcal{T}:\boldsymbol{k}\mapsto -D_{2}\cdot\boldsymbol{k} = \boldsymbol{k}$. The action of the $\mathcal{C}_2\mathcal{T}$ symmetry on the (Zak) Bloch basis is
\begin{equation}
\begin{aligned}
    ^{\mathcal{C}_2\mathcal{T}}\vert \boldsymbol{\varphi}^Z,\boldsymbol{k}\rangle &= \vert \boldsymbol{\varphi}^Z,-D_2\cdot \boldsymbol{k}\rangle\cdot  U_{2'} \mathcal{K}
    \,, \\
    &=  \vert \boldsymbol{\varphi}^Z,\boldsymbol{k}\rangle\cdot  U_{2'} \mathcal{K}
    \,,
\end{aligned}
\end{equation}
with $\mathcal{K}$ the complex conjugation operator (acting on the right) and $U_{2'}$ the unitary matrix that captures the effect of $\mathcal{C}_2\mathcal{T}$ on the sublattice and orbital degrees of freedom. 

The Bloch Hamiltonian matrix must then satisfy the condition
\begin{equation}
    U_{2'} \cdot H^Z(\boldsymbol{k})^*\cdot U_{2'}^{\dagger} = H^Z(\boldsymbol{k}) \,.
\end{equation}
The condition $[\mathcal{C}_2\mathcal{T}]^2=\mathbb{1}$ then implies for the representation of $\mathcal{C}_2\mathcal{T}$ in the Zak basis
\begin{equation}
    U_{2'} \cdot U_{2'}^* = \mathbb{1}_{M\times M}\,.
\end{equation}
Since $U_{2'}$ is unitary, the reality condition implies that it must also be symmetric and it admits a Takagi factorization.

\subsubsection{$\mathcal{C}_2\mathcal{T}$ Representation and Shift (Phase) Matrix for Separated $\mathcal{C}_2$-symmetric and Non-$\mathcal{C}_2$-symmetric WPs}

Assuming the ordering of the sublattice and orbital degrees of freedom of Eqs.\;(\ref{eq_indices}), the representation of $\mathcal{C}_2\mathcal{T}$ in the subspace $(\rho,l)$ of a WP $\rho$ and orbital $l$ takes the two following forms. In the case of a $\mathcal{C}_2$-symmetric WP $\rho^*$, we have
\begin{equation}
    U_{2'}^{(\rho^*,l)}
     =  \mathbb{1}_{n_{\rho^*}\times n_{\rho^*}}
    \otimes U_{2',l} \,,
\end{equation}
and, in the case of a non-$\mathcal{C}_2$-symmetric WP $\rho$, we have
\begin{equation}
    U_{2'}^{(\rho,l)}
     =  \mathbb{1}_{n_{\rho}/2\times n_{\rho}/2}
    \otimes \sigma_x \otimes U_{2',l} \,,
\end{equation}
where $U_{2',l}$ is the unitary matrix of rank $n_{l}$ acting on the orbital subspace. The global representation is then given by the direct sum (i.e. catenation of diagonal blocks) over all the WPs, i.e.
\begin{equation}
    U_{2'} = \bigoplus\limits_{\rho} \bigoplus\limits_{l\in J_{\rho}}
    \bigoplus\limits_{\beta =1}^{ n_{\rho,l} }
    U_{2'}^{(\rho^*,l^{\beta})}\,.
\end{equation}

The shift matrix in the subspace $(\rho^*,l)$ of a $\mathcal{C}_2$-symmetric WP is 
\begin{equation}
    \left[V^Z(\boldsymbol{G})\right]_{(\rho^*,l)} = \text{diag} \left( 
        e^{\text{i} \boldsymbol{G}\cdot \boldsymbol{r}_{\rho^*_1}} , \dots ,
        e^{\text{i} \boldsymbol{G}\cdot \boldsymbol{r}_{\rho^*_{n_{\rho^*}}} }
    \right)\otimes \mathbb{1}_{n_l\times n_l} \,,
\end{equation}
while for a non-symmetric WP $\rho$ it is
\begin{equation}
\left[V^Z(\boldsymbol{G})\right]_{(\rho,l)} = \bigoplus\limits_{j=1}^{n_{\rho}/2} \left[\text{diag}\left( e^{\text{i} \boldsymbol{G}\cdot \boldsymbol{r}_{\rho_{2j}}} , 
e^{-\text{i} \boldsymbol{G}\cdot \boldsymbol{r}_{\rho_{2j}}}
\right) \otimes \mathbb{1}_{n_l\times n_l} \right] .
\end{equation}

We now verify that the $\mathcal{C}_2\mathcal{T}$ transformed basis at a shifted point $\boldsymbol{k}+\boldsymbol{G}$, i.e. 
\begin{equation}
\begin{aligned}
    ^{\mathcal{C}_2\mathcal{T}}\vert \boldsymbol{\varphi}^Z,\boldsymbol{k}+\boldsymbol{G}\rangle &= \vert \boldsymbol{\varphi}^Z,-D_2\cdot (\boldsymbol{k}+\boldsymbol{G})\rangle \cdot U_{2'} \mathcal{K} \\
    & = \vert \boldsymbol{\varphi}^Z,\boldsymbol{k}+\boldsymbol{G}\rangle \cdot U_{2'} \mathcal{K} \,,
\end{aligned}  
\end{equation}
must match with 
\begin{equation}
\begin{aligned}
    &^{\mathcal{C}_2\mathcal{T}}\vert \boldsymbol{\varphi}^Z,\boldsymbol{k}+\boldsymbol{G}\rangle = \left.^{\mathcal{C}_2\mathcal{T}}\vert \boldsymbol{\varphi}^Z,\boldsymbol{k}\rangle\right. \cdot V^Z(\boldsymbol{G}) \\
    &= \vert \boldsymbol{\varphi}^Z,-D_2 \boldsymbol{k}\rangle \cdot U_{2'} \cdot V^Z(-\boldsymbol{G}) \mathcal{K}\\
    &= \vert \boldsymbol{\varphi}^Z,-D_2 \boldsymbol{k}-D_2\boldsymbol{G}\rangle \cdot \left[V^Z(D_2 \boldsymbol{G}) \cdot U_{2'} \cdot V^Z(-\boldsymbol{G}) \right]\mathcal{K}\\
    &= \vert \boldsymbol{\varphi}^Z, \boldsymbol{k}+\boldsymbol{G}\rangle \cdot \left[V^Z(- \boldsymbol{G}) \cdot U_{2'} \cdot V^Z(-\boldsymbol{G}) \right]\mathcal{K}
    \,.
\end{aligned}  
\end{equation}
That is, the following relation must be satisfied 
\begin{equation}
    U_{2'} = V^Z(- \boldsymbol{G}) \cdot U_{2'} \cdot V^Z(-\boldsymbol{G})\,.
\end{equation}

It can be easily verified that the above consistency relation is satisfied by the $\mathcal{C}_2$-center subspaces if and only if
\begin{equation}
    e^{\text{i} \boldsymbol{G}\cdot \left( D_2\boldsymbol{r}_{\rho^*_j} - \boldsymbol{r}_{\rho^*_j} \right)} = 1\,,
\end{equation}
 or, equivalently, 
\begin{equation}
\begin{aligned}
     D_2\boldsymbol{r}_{\rho^*_j} - \boldsymbol{r}_{\rho^*_j} &= \boldsymbol{R}_{j} \in \text{BL}\,,\\
     \Leftrightarrow\quad 
     \boldsymbol{r}_{\rho^*_j} &= -\dfrac{1}{2}\boldsymbol{R}_{j} = -\dfrac{n_{j,1}}{2} \boldsymbol{a}_1
     -\dfrac{n_{j,2}}{2} \boldsymbol{a}_2\,,
\end{aligned}
\end{equation}
for all $j=1,\dots, n_{\rho^*}$ and $n_{j,1},n_{j,2}\in \mathbb{Z}$, and for all $\boldsymbol{G}$ in the reciprocal Bravais lattice. Thus, the phase factors entering the shift matrix ($\left[V^Z(\boldsymbol{G})\right]_{(\rho^*,l)}$) are all of the form, with $m_1,m_2\in \mathbb{Z}$,
\begin{equation}
\label{eq_c2sym_shift}
\begin{aligned}
    e^{\text{i} \boldsymbol{G}\cdot \boldsymbol{r}_{\rho^*_j}} &= 
    e^{-\text{i} (m_1 \boldsymbol{b}_1+m_2 \boldsymbol{b}_2)\cdot \left(\dfrac{n_{j,1}}{2} \boldsymbol{a}_1
     +\dfrac{n_{j,2}}{2} \boldsymbol{a}_2\right)} \\
     &= e^{-\text{i} (m_1 n_{j,1}+m_2 n_{j,2}) \pi} \in \{+1,-1\} \,.
\end{aligned}
\end{equation}
We thus conclude that the phase matrix in the $\mathcal{C}_2$-symmetric subspace is a diagonal matrix of $\pm1$'s.

For non-$\mathcal{C}_2$-symmetric subspaces, the consistency relation is satisfied if and only if 
\begin{equation}
    e^{\text{i} \boldsymbol{G}\cdot \left( D_2 \boldsymbol{r}_{\rho_{2j}} - \boldsymbol{r}_{\rho_{2j-1}} \right) } =1\,,
\end{equation}
for all $j=1,\dots, n_{\rho}/2$, and for all reciprocal Bravais lattice vector $\boldsymbol{G}$. Equivalently, that is 
\begin{equation}
    D_2 \boldsymbol{r}_{\rho_{2j}} - \boldsymbol{r}_{\rho_{2j-1}} \in \text{BL}\,,
\end{equation}
or $[D_2 \boldsymbol{r}\rho_{2j}] = [\boldsymbol{r}_{\rho_{2j-1}}]$, again for $j=1,2\dots n_{\rho}/2$. Since $D_2\boldsymbol{r} = -\boldsymbol{r}$ in a 2D system perpendicular to the $\mathcal{C}_2$-axis, the shift matrix for the non-$\mathcal{C}_2$-centers in the Zak-Bloch basis has the form
\begin{equation}
\begin{aligned}
    [V^Z(\boldsymbol{G})]_{\rho_{2j-1},\rho_{2j}} &= 
    \text{diag}
    \left( e^{\text{i} \boldsymbol{G}\cdot \boldsymbol{r}_{\rho_{2j-1}}} , 
    e^{\text{i} \boldsymbol{G}\cdot \boldsymbol{r}_{\rho_{2j}}} \right)\,,\\
    &= \text{diag}
    \left( e^{\text{i} \boldsymbol{G}\cdot \boldsymbol{r}_{\rho_{2j-1}}} , 
    e^{-\text{i} \boldsymbol{G}\cdot \boldsymbol{r}_{\rho_{2j-1}}} \right)\,.
\end{aligned}
\end{equation}
Contrary to the subspace of $C_2$-centers, the shift matrix is complex for some reciprocal vectors $\boldsymbol{G}$.

\subsubsection{Takagi Factorization}\label{sec_takagifact}

Being symmetric, the $\mathcal{C}_2\mathcal{T}$ representation matrix $U_{2'}$ admits a Takagi factorization $U_{2'} = U_{\text{TF}} \cdot U_{\text{TF}}^T$. In general, it can be determined through the singular value decomposition $U_{2'} = U_{\text{svd}}\cdot \Sigma \cdot V_{\text{svd}}$, from which we get $U_{\text{TF}} = U_{\text{svd}} \cdot \sqrt{U_{\text{svd}}^{\dagger}\cdot V_{\text{svd}}^*}$ \cite{CHEBOTAREV2014380}. Since $U_{2'}$ is also unitary, one easily verifies that we can take
\begin{equation}
\label{eq_TF_effansatz}
    U_{\text{TF}} = \sqrt{U_{2'}} \,,
\end{equation}
[see Eq.\;(\ref{eq_TF_C2Tidentity})] even though it is not directly obvious that the above expression based on singular value decomposition reduces to this simple form.

We then define the change of basis
\begin{equation}
    \vert \widetilde{\boldsymbol{\varphi}},\boldsymbol{k} \rangle = \vert \boldsymbol{\varphi}^Z,\boldsymbol{k}\rangle  \cdot U_{\text{TF}} = \vert \boldsymbol{\varphi}^Z,\boldsymbol{k}\rangle  \cdot \sqrt{U_{2'}} \,,
\end{equation}
in which the Bloch Hamiltonian matrix transforms to
\begin{equation}
    \widetilde{H}(\boldsymbol{k}) = \sqrt{U_{2'}} \cdot H^Z(\boldsymbol{k}) \cdot \sqrt{U_{2'}^*} \,.
\end{equation}
The representation of $C_2T$ in the new basis is then
\begin{equation}
\label{eq_TF_C2Tidentity}
\begin{aligned}
    ^{C_2T}\vert \widetilde{\boldsymbol{\varphi}},\boldsymbol{k} \rangle &= \left.^{C_2T}\vert \boldsymbol{\varphi}^Z,\boldsymbol{k}\rangle\right.  \cdot \sqrt{U_{2'}} \,,\\
    &= \vert \widetilde{\boldsymbol{\varphi}},\boldsymbol{k}\rangle \cdot \left[ \sqrt{U_{2'}^*} \cdot U_{2'} \cdot \sqrt{U_{2'}^*} \right] \mathcal{K} \,,\\
    &= \vert \widetilde{\boldsymbol{\varphi}},\boldsymbol{k}\rangle \cdot \left[ \sqrt{U_{2'}^{\dagger}} \cdot U_{2'} \cdot \sqrt{U_{2'}^{\dagger}} \right] \mathcal{K} \,,\\
    &= \vert \widetilde{\boldsymbol{\varphi}},\boldsymbol{k}\rangle \; \mathcal{K}\,,
\end{aligned}
\end{equation}
i.e.~the unitary part of $\mathcal{C}_2\mathcal{T}$ representation in the new basis is the identity matrix [it is critical for this that $U_{2'}$ is unitary, such that it commutes with its conjugate transpose], and the condition on the transformed Bloch Hamiltonian matrix reduces to
\begin{equation}
    \widetilde{H}(\boldsymbol{k})^* = \widetilde{H}(\boldsymbol{k})\,,
\end{equation}
and must thus be real symmetric. 

Shifting the new basis by a full reciprocal lattice vector $\boldsymbol{G}$, we get
\begin{equation}
\left\{
\begin{aligned}
    \vert \widetilde{\boldsymbol{\varphi}},\boldsymbol{k}+ \boldsymbol{G} \rangle &= 
    \vert \widetilde{\boldsymbol{\varphi}},\boldsymbol{k} \rangle \cdot \widetilde{V}(\boldsymbol{G})
    \,,\\
    \widetilde{V}(\boldsymbol{G}) &= 
    \sqrt{U_{2'}^*} \cdot V^Z(\boldsymbol{G})\cdot \sqrt{U_{2'}} \,.
\end{aligned}\right.
\end{equation}
Then, assuming the above separation between $\mathcal{C}_2$-centers and non-$\mathcal{C}_2$-centers, and noting that the square root and complex conjugation of a tensor product of matrices is given by the tensor product of the square root and complex conjugation of the matrices, i.e.
\begin{equation}
\label{eq_blockdiagonal_U2}
\left\{
    \begin{aligned}
        \sqrt{ U_{2'}^{(\rho^*,l)} } &=  \mathbb{1}_{n_{\rho^*}\times n_{\rho^*}}
    \otimes \sqrt{U_{2',l}} \,,\\
        \sqrt{ U_{2'}^{(\rho,l)} } &=  \mathbb{1}_{n_{\rho}/2\times n_{\rho}/2}
    \otimes \sqrt{\sigma_x} \otimes \sqrt{U_{2',l}} \\
    &= \mathbb{1}_{n_{\rho}/2\times n_{\rho}/2}
    \otimes \frac{1}{2}\left[\substack{1+\text{i} ~ 1-\text{i}\\
    1-\text{i} ~ 1+\text{i} }\right] \otimes \sqrt{U_{2',l}}\,,
    \end{aligned}\right.
\end{equation}
we obtain for the transformed shift matrix $\widetilde{V}(\boldsymbol{G})$ that
\begin{equation}
\label{eq_TF_shift1}
\left\{
\begin{aligned}
    &\widetilde{V}^{{(\rho^*,l)}}(\boldsymbol{G}) = V^{Z}_{(\rho^*,l)}(\boldsymbol{G})\in \mathbb{R}^{n_{\rho^*} n_l} \times \mathbb{R}^{n_{\rho^*} n_l}\,,\\
    &\widetilde{V}^{(\rho,l)}(\boldsymbol{G}) =\\
    & \bigoplus\limits_{j=1}^{n_{\rho}/2}
    \left[
        \begin{array}{cc}
             \cos \left(\boldsymbol{G}\cdot \boldsymbol{r}_{\rho_{2j-1}}\right) & \sin \left(\boldsymbol{G}\cdot \boldsymbol{r}_{\rho_{2j-1}}\right) \\
            -\sin \left(\boldsymbol{G}\cdot \boldsymbol{r}_{\rho_{2j-1}}\right)  & \cos \left(\boldsymbol{G}\cdot \boldsymbol{r}_{\rho_{2j-1}}\right)
        \end{array}
    \right]\otimes \mathbb{1}_{n_l\times n_l},
\end{aligned}\right.
\end{equation}
where we have used that $U_{2',l}^* = U_{2',l}^{\dagger} = U_{2',l}^{-1}$. While the shift matrices are unchanged for $\mathcal{C}_2$-symmetric subspaces, i.e.~diagonal matrix of $\pm1$ (Eq.\;(\ref{eq_c2sym_shift})), they become real for the non-$\mathcal{C}_2$-symmetric subspaces as well, albeit not diagonal. Again, the obtained real Hamiltonian matrix is not periodic under reciprocal lattice translations in general. Next, we discuss the consequences of this on the Zak phases associated to non-$\mathcal{C}_2$-symmetric atomic sites. In Section ~\ref{sec_maxper}, we will derive the change of basis required to obtain a \textit{maximally periodic} real Bloch Hamiltonian matrix, i.e.~such that the shift matrix becomes maximally diagonal without altering the topological information and physical interpretation captured by the Zak phases. Given the reality of the Hamiltonian, by definition the entries of the shift matrix in this case will be $\pm1$ in general for all atomic orbitals.

\subsection{1D  Topology Indicated by Maximally Periodic Real Bloch Hamiltonians}\label{sec_Zakphase_Z2}

We have derived above that the reality condition guarantees the existence of a change of basis, obtained through a Takagi factorization, in which the Bloch Hamiltonian matrix and its associated shift matrices $\widetilde{V}(\boldsymbol{G})$ are all real for all $\boldsymbol{G}\in \text{BL}^*$ (the reciprocal Bravais lattice). A direct consequence of the real Bloch form is that Wilson loops now define $\mathsf{O}(p_{\nu})$ orthonormal matrices, see Eq.\;(\ref{eq_Zak_phase_definition}). The Zak phase factors are thus restricted to 
\begin{equation}
    e^{\text{i} \gamma_{\nu}[\boldsymbol{k}_0+\boldsymbol{G}\leftarrow \boldsymbol{k}_0]} = \det \widetilde{\mathcal{W}}_{\nu, [\boldsymbol{k}_0+\boldsymbol{G}\leftarrow \boldsymbol{k}_0]} \in \{+1,-1\}\,,
\end{equation}
that is, the reality condition implies the $\mathbb{Z}_2$ quantization of Zak phases (as long as the band-subspace has any non-$\mathcal{C}_2$-symmetric sites included in corresponding pairs) 
\begin{equation}
    \gamma_{\nu}[\boldsymbol{k}_0+\boldsymbol{G}\leftarrow \boldsymbol{k}_0] \in \{0,\pi\} \mod \,2\pi\,, 
\end{equation}
given a $\nu$-th band subspace (of rank $p_{\nu}$) isolated from the other bands. The Zak phase captures the orientability of the one-dimensional band structure, that is, whether the orientation of the $p_{\nu}$-dimensional frame of Bloch eigenvectors $\left\langle 
u^Z_{\nu+1}(\boldsymbol{k}), \cdots, u^Z_{\nu+p_{\nu}}(\boldsymbol{k}) \right\rangle $ can be chosen continuously over the whole 1D reciprocal direction (i.e.~the {\it whole} line joining $\boldsymbol{k}_0$ and $\boldsymbol{k}_0+\boldsymbol{G}$), in which case the Zak phase is zero (modulo $2\pi$). A $\pi$ Zak phase implies that an odd number of Bloch eigenvectors, {\it defined in the periodic gauge} (Sec.\;\ref{sec_periodicity_Blochop}), undergo a $\pi$-phase jump from one Brillouin zone cell $[\boldsymbol{k}_0,\boldsymbol{k}_0+\boldsymbol{G}]$ to the next $[\boldsymbol{k}_0+\boldsymbol{G},\boldsymbol{k}_0+2\boldsymbol{G}]$, in which case we say that the band structure is non-orientable. 

Importantly, we find that the real Wilson loop of atomic flag limits [Sec.\;\ref{secAppx:AtomicFlagLim}], when obtained from the above transformed basis ($\varphi \rightarrow \widetilde{\varphi}$), is not diagonal whenever there are non-$\mathcal{C}_2$-symmetric WPs, contrary to the Wilson loop expressed in the Zak-Bloch basis [Eq.\;(\ref{eq_Zak_flag_atomic_limit})]. Indeed, in an atomic flag limit we get 
\begin{equation}
    \widetilde{\mathcal{W}}_{ [\boldsymbol{k}_0+\boldsymbol{G}\leftarrow \boldsymbol{k}_0]} = 
    \widetilde{V}(\boldsymbol{G})
    \,,
\end{equation}
with $\widetilde{V}(\boldsymbol{G})$ given in Eq.\;(\ref{eq_TF_shift1}). While the Wilson loop eigenvalues of a non-$\mathcal{C}_2$-symmetric sublattice pair, $\{\boldsymbol{r}_{2j-1},\boldsymbol{r}_{2j}\}$, are complex, i.e. $\{e^{\text{i} \boldsymbol{G}\cdot \boldsymbol{r}_{2j-1}},e^{ \text{i} \boldsymbol{G}\cdot \boldsymbol{r}_{2j}}\}$, the Zak phase of this atomic subspace is zero since, given $\boldsymbol{r}_{2j}=-\boldsymbol{r}_{2j-1}$, we have
\begin{equation}
    e^{\text{i} \gamma_{\{\rho_{2j-1},\rho_{2j}\}}[\boldsymbol{G}]} = e^{\text{i} \boldsymbol{G}\cdot \boldsymbol{r}_{2j-1}} e^{\text{i} \boldsymbol{G}\cdot \boldsymbol{r}_{2j}}=1\,,
\end{equation} 
and the determinant of the Wilson loop is unaffected by the change of basis during the Takagi factorization procedure. Clearly however, the Zak phase associated to an individual non-$\mathcal{C}_2$-symmetric atomic orbital is not $\mathbb{Z}_2$ quantized in general. We will show in the next section that this pair-wise cancellation of the Zak phase, in the atomic flag limit, associated to non-$C_2$-symmetric sublattice sites, permits us to change the Fourier phase factor of their associated Bloch basis states, making the transformed Bloch Hamiltonian matrix periodic over a larger fraction of the sublattice degrees of freedom.

While Zak phases capture the periodicity of Bloch eigenvectors along commensurate directions of the reciprocal space, i.e.~their orientability, there is {\it a priori} no direct relation between the orientability of band structures and the periodicity of the real Bloch Hamiltonian matrix $\widetilde{H}(\boldsymbol{k})$. After all, we saw in Section \ref{sec_periodicity_Blochop} that the non-periodicity of the Bloch Hamiltonian matrix is associated to the choice of the Zak basis as the Bloch basis. This leads to the question of {\it whether one can get rid of the non-periodicity of the Bloch Hamiltonian matrix associated to specific atomic degrees of freedom, while preserving the topological data carried by the Bloch eigenvectors, that is, preserving the Zak phases} (provided that the band-subspaces are partitioned to always include non-$\mathcal{C}_2$-symmetric sites in corresponding pairs). Indeed, the non-orientability of bands, that is also related to the charge anomaly of topological edge states in Su-Schrieffer-Heeger (SSH) phases, should not depend on the chosen Bloch basis (under the assumption that the chosen origin of the Bravais lattice is not changed). 

We show in the following that the non-periodicity of the Bloch Hamiltonian matrix $ \widetilde{H}(\boldsymbol{k})$ associated to the non-$\mathcal{C}_2$-symmetric atomic degrees of freedom can be removed through a change of the Bloch basis, while the atomic flag limit  Zak phases of all band-subspaces that include non-$\mathcal{C}_2$-symmetric atomic orbitals in corresponding pairs are preserved and, incidentally, so is the principle of the SSH bulk-boundary-correspondence. We refer to such form of the Bloch Hamiltonian Matrix as being \textit{maximally periodic} since the shift matrix will be diagonal with $\pm1$ entries as demonstrated below.

\subsubsection{Maximal Periodicity and Altered  1D Topology of Atomic Flag Limits}\label{sec_maxper}

We first address the question of the existence of a change of basis that brings the Bloch Hamiltonian matrix to the most periodic form possible, while preserving its real valuedness. 

It readily follows from the definition of the Zak-Bloch basis Eq.\;(\ref{eq_Bloch_basis_Zak}) that the change of basis 
\begin{equation}\label{eq:fullyperiodic_basis}
    \vert \varphi^Z_{\alpha_i},\boldsymbol{k} \rangle = \vert \overline{\varphi}_{\alpha_i},\boldsymbol{k} \rangle \;e^{ \text{i} \boldsymbol{k}\cdot \boldsymbol{r}_{\alpha_i}}\,,
\end{equation}
from which we define the matrix of phase factors
\begin{equation}
    [G(\boldsymbol{k})]_{\alpha_i \alpha_j} = e^{ \text{i} \boldsymbol{k}\cdot \boldsymbol{r}_{\alpha_i}} \delta_{ij}\,,
\end{equation}
gives a periodic Bloch Hamiltonian matrix
\begin{equation}
    \overline{H}(\boldsymbol{k}) = G(\boldsymbol{k})\cdot H^Z(\boldsymbol{k}) \cdot G(-\boldsymbol{k})\,,
\end{equation}
since 
\begin{equation}
\begin{aligned}
    \overline{H}(\boldsymbol{k}+\boldsymbol{G}) &= G(\boldsymbol{k}+\boldsymbol{G}) H^Z(\boldsymbol{k}+\boldsymbol{G}) G(-\boldsymbol{k}-\boldsymbol{G}) \,,\\
    &= G(\boldsymbol{k})V^Z(\boldsymbol{G}) H^Z(\boldsymbol{k}+\boldsymbol{G}) 
    V^Z(-\boldsymbol{G})G(-\boldsymbol{k}) \,,\\
    &= G(\boldsymbol{k}) H^Z(\boldsymbol{k}) 
    G(-\boldsymbol{k}) = \overline{H}(\boldsymbol{k})\,.
\end{aligned}
\end{equation}
Crucially, one readily finds for atomic flag limits that this change of basis, if done for all orbitals, trivializes all the $\pi$-Zak phases associated with the atomic orbitals localized at $\mathcal{C}_2$-symmetric WPs on the unit cell boundary. Furthermore, while this change of basis brings periodicity in all reciprocal directions, it also changes the representation of $\mathcal{C}_2\mathcal{T}$ and, following, the real valuedness of the Bloch Hamiltonian matrix. Take the kagome lattice, for instance: the above change of basis removes all $\pi$-Zak phases of atomic flag limits, even though the atomic orbitals are all off-centered, occupying the $\mathcal{C}_2$-symmetric WP $3c$, as pointed out above, and it takes the initially real Hamiltonian in the Zak-Bloch basis to a complex valued expression.   

One thus needs a general criterion as to which atomic degrees of freedom the above change of basis (a change of Fourier transform's phase factor) makes the real-valued Bloch Hamiltonian maximally periodic without affecting the 1D topological invariants ($\pi$-Zak phases) and their physical interpretation (unit cell off-centering). 

\subsubsection{General Criterion To Obtain Maximally Periodic Bloch Basis with Preserved 1D Topology}

Let us again assume the separation of the sublattice degrees of freedom from the others (e.g. the orbitals), as well as their separation into $\mathcal{C}_2$-centers ($\rho^*$) and non-$\mathcal{C}_2$-centers ($\rho$). We first rewrite the above change of basis for the atomic subspace associated to a sublattice site $\boldsymbol{r}_{\rho^*_j}$ belonging to a $\mathcal{C}_2$-symmetric WP $\rho^*$ and the $l$-th orbital components populating the site, as well as for the atomic subspace associated to a pair of sublattice sites $\{\boldsymbol{r}_{\rho_{2j-1}},\boldsymbol{r}_{\rho_{2j}}\}$ belonging to a non-$\mathcal{C}_2$-symmetric WP $\rho$ and the $l$-th orbital components populating the pair of sites. Writing the Bloch basis associated to a $(\rho_j,l)$-subspace as
\begin{equation}
\vert \boldsymbol{\varphi}^Z_{(\rho_j,l)} ,\boldsymbol{k}\rangle  = \left(
            \vert \varphi^Z_{\rho_j,l_1} ,\boldsymbol{k}\rangle~\cdots 
            \vert \varphi^Z_{\rho_j,l_{n_l}} ,\boldsymbol{k}\rangle
        \right)\,,
\end{equation}
the above change of basis reads, for a $\mathcal{C}_2$-center,
\begin{equation}
\left\{
    \begin{aligned}
        \vert \boldsymbol{\varphi}^Z_{(\rho^*_j,l)} ,\boldsymbol{k}\rangle  &=  \vert \overline{\boldsymbol{\varphi}}_{(\rho^*_j,l)} ,\boldsymbol{k}\rangle \cdot G^{(\rho^*_j,l)}(\boldsymbol{k})  \,,\\
       G^{(\rho^*_j,l)}(\boldsymbol{k}) &= e^{\text{i}  \boldsymbol{k}\cdot \boldsymbol{r}_{\rho^*_j}}\, \mathbb{1}_{n_l\times n_l}\,,
    \end{aligned}\right.
\end{equation}
and for a pair of non-$\mathcal{C}_2$-symmetric sites,
\begin{equation}
\left\{
\begin{aligned}
     &\left(\vert \boldsymbol{\varphi}^Z_{(\rho_{2j-1},l)} ,\boldsymbol{k}\rangle ~
        \vert \boldsymbol{\varphi}^Z_{(\rho_{2j},l)} ,\boldsymbol{k}\rangle
        \right)  = \\
        &\quad \left(\vert \overline{\boldsymbol{\varphi}}_{(\rho_{2j-1},l)} ,\boldsymbol{k}\rangle ~
        \overline{\vert \boldsymbol{\varphi}}_{(\rho_{2j},l)} ,\boldsymbol{k}\rangle
        \right)\cdot G^{(\rho_{j},l)}(\boldsymbol{k}) \,,\\
    & G^{(\rho_{j},l)}(\boldsymbol{k}) = 
    \text{diag}\left(  
        e^{\text{i}  \boldsymbol{k}\cdot \boldsymbol{r}_{\rho_{2j-1}}}\, ,\,
        e^{-\text{i}  \boldsymbol{k}\cdot \boldsymbol{r}_{\rho_{2j-1}}}
    \right)
    \otimes \mathbb{1}_{n_l\times n_l} \,.
\end{aligned}\right.
\end{equation} 
In order to keep notation compact, we will now simply write 
\begin{equation}
    \vert \boldsymbol{\varphi}^Z_{(\rho_j,l)} ,\boldsymbol{k}\rangle  = \vert \overline{\boldsymbol{\varphi}}_{(\rho_j,l)} ,\boldsymbol{k}\rangle \cdot G^{(\rho_j,l)}(\boldsymbol{k}) \,,
\end{equation}
and ask the reader to rewrite it into the above expressions for the different cases.

The representation of $\mathcal{C}_2\mathcal{T}$ in the new basis is then
\begin{equation}
\left\{
    \begin{aligned}
        \left.^{\mathcal{C}_2\mathcal{T}}\right.\vert \overline{\boldsymbol{\varphi}}_{(\rho_j,l)},\boldsymbol{k} \rangle &= \vert \overline{\boldsymbol{\varphi}}_{(\rho_j,l)} , \boldsymbol{k} \rangle \cdot \overline{U}_{2'}^{(\rho_j,l)}(\boldsymbol{k}) \,\mathcal{K} \,,\\
          \overline{U}_{2'}^{(\rho_j,l)}(\boldsymbol{k}) &= 
          G^{(\rho_j,l)}(\boldsymbol{k}) \cdot U_{2'}^{(\rho_j,l)} \cdot G^{(\rho_j,l)}(\boldsymbol{k})  \,,
    \end{aligned}\right.
\end{equation}
and the shift by a reciprocal vector is now simply
\begin{equation}
    \vert \overline{\boldsymbol{\varphi}}_{(\rho_j,l)},\boldsymbol{k}+\boldsymbol{G} \rangle =
        \vert \overline{\boldsymbol{\varphi}}_{(\rho_j,l)},\boldsymbol{k} \rangle\,.
\end{equation}

Clearly, the intra-sublattice-site terms of the Bloch Hamiltonian matrix, now written in the new basis, associated to the subspaces $(\rho_j,l)$ that underwent the above change of basis are now periodic under any reciprocal Bravais translation $\boldsymbol{G}$, i.e. (compactly written)
\begin{equation}
    [\overline{H}(\boldsymbol{k}+\boldsymbol{G})]_{(\rho_j,l),(\rho_j,l)} = [\overline{H}(\boldsymbol{k})]_{(\rho_j,l),(\rho_j,l)} \,. 
\end{equation}

The Bloch Hamiltonian matrix may still be complex, i.e. whenever there are atomic subspaces $(\rho_j,l)$ for which the $\mathcal{C}_2\mathcal{T}$ representation is not simply the complex conjugation $\mathcal{K}$ (i.e. if not all $\overline{U}_{2'}^{(\rho_j,l)}(\boldsymbol{k})$ are the identity matrix). The Takagi factorization of the $\mathcal{C}_2\mathcal{T}$ representation then provides the change of basis that makes the Bloch Hamiltonian real. From the result of Section \ref{sec_takagifact}, we have  (again written compactly)
\begin{equation}
\left\{
    \begin{aligned} 
    \vert \widetilde{\overline{\boldsymbol{\varphi}}}_{(\rho_j,l)},\boldsymbol{k} \rangle &= \vert \overline{\boldsymbol{\varphi}}_{(\rho_j,l)} , \boldsymbol{k} \rangle \cdot \overline{U}_{TF}^{(\rho_j,l)}(\boldsymbol{k})^{\dagger}  \,,\\
    \overline{U}_{TF}^{(\rho_j,l)}(\boldsymbol{k}) &= 
    \sqrt{\overline{U}_{2'}^{(\rho_j,l)}(\boldsymbol{k})} \,,
    \end{aligned}\right.
\end{equation}
such that the $\mathcal{C}_2\mathcal{T}$ representation is just complex conjugation
\begin{equation}
    \left.^{\mathcal{C}_2\mathcal{T}}\right.\vert \widetilde{\overline{\boldsymbol{\varphi}}}_{(\rho_j,l)},\boldsymbol{k} \rangle = \vert \widetilde{\overline{\boldsymbol{\varphi}}}_{(\rho_j,l)},\boldsymbol{k} \rangle\;\mathcal{K} \,.
\end{equation}
A shift by a reciprocal Bravais vector is finally given by
\begin{equation}
    \vert \widetilde{\overline{\boldsymbol{\varphi}}}_{(\rho_j,l)},\boldsymbol{k}+\boldsymbol{G} \rangle = 
    \vert \widetilde{\overline{\boldsymbol{\varphi}}}_{(\rho_j,l)},\boldsymbol{k}\rangle \cdot \widetilde{\overline{V}}^{(\rho_j,l)}_{\boldsymbol{k}}(\boldsymbol{G}) \,,
\end{equation}
with the transformed shift matrix
\begin{equation}
\label{eq_shift_criterium}
    \begin{aligned} 
    & \widetilde{\overline{V}}^{(\rho_j,l)}_{\boldsymbol{k}}(\boldsymbol{G}) = \overline{U}_{TF}^{(\rho_j,l)}(\boldsymbol{k})^{\dagger} \cdot \overline{U}_{TF}^{(\rho_j,l)}(\boldsymbol{k}+\boldsymbol{G})\,,\\
    &= \sqrt{\overline{U}_{2'}^{(\rho_j,l)}(\boldsymbol{k})^{\dagger} } \cdot \sqrt{\overline{U}_{2'}^{(\rho_j,l)}(\boldsymbol{k}+\boldsymbol{G})}\,,\\
    &=  \sqrt{ \overline{U}_{2'}^{(\rho_j,l)}(\boldsymbol{k})^* } \cdot \sqrt{V^Z_{(\rho_j,l)}(\boldsymbol{G})\cdot \overline{U}_{2'}^{(\rho_j,l)}(\boldsymbol{k}) \cdot V^Z_{(\rho_j,l)}(\boldsymbol{G}) } \,,\\
    &=  \sqrt{
    G^{(\rho_j,l)}(-\boldsymbol{k})\cdot 
    U_{2'}^{(\rho_j,l)*} \cdot  G^{(\rho_j,l)}(-\boldsymbol{k}) 
    } \quad \cdot \, \\
    & \sqrt{V^Z_{(\rho_j,l)}(\boldsymbol{G})\cdot 
    G^{(\rho_j,l)}(\boldsymbol{k})\cdot 
    U_{2'}^{(\rho_j,l)} \cdot  
    G^{(\rho_j,l)}(\boldsymbol{k}) \cdot V^Z_{(\rho_j,l)}(\boldsymbol{G}) }
    \end{aligned}
\end{equation}
which, in the last line, is given only in terms of the $\mathcal{C}_2\mathcal{T}$ representation, $U_{2'}$, the phase-factor matrix, $G(\boldsymbol{k})$, and shift matrix, $V^Z(\boldsymbol{G})$, all in the initial Zak-Bloch basis $\vert \boldsymbol{\varphi}^Z_{(\rho_j,l)},\boldsymbol{k}\rangle$. 

We can now formulate the general criterion: {\it the non-periodic Zak-Bloch basis of the atomic degrees of freedom $(\rho_j,l)$ can be changed to a (fully) periodic basis ($\vert \boldsymbol{\varphi}^Z_{(\rho_j,l)} , \boldsymbol{k}\rangle \mapsto \vert \overline{\boldsymbol{\varphi}}_{(\rho_j,l)} , \boldsymbol{k}\rangle $) and then be mapped to the basis that makes the Bloch Hamiltonian real valued, through the Takagi factorization of the $\mathcal{C}_2\mathcal{T}$ representation ($\vert \overline{\boldsymbol{\varphi}}_{(\rho_j,l)} , \boldsymbol{k}\rangle \mapsto \vert \widetilde{\overline{\boldsymbol{\varphi}}}_{(\rho_j,l)} , \boldsymbol{k}\rangle$, as above), while preserving the 1D topology of band-subspaces (provided non-$\mathcal{C}_2$-symmetric sites are included in corresponding pairs) in atomic flag limits, whenever the shift matrix in the final basis, defined in Eq.\;(\ref{eq_shift_criterium}), is identity for the $(\rho,l)$ non-$\mathcal{C}_2$-symmetric WPs subspace, or is equal to $V^{Z}_{(\rho^*,l)}$ for the $\mathcal{C}_2$ symmetric WPs subspace, that is when 
\begin{equation}
\begin{aligned}
    &\widetilde{\overline{V}}^{(\rho,l)}_{\boldsymbol{k}}(\boldsymbol{G}) = \mathbb{1} \,,\\
    \text{or} \quad\quad & \widetilde{\overline{V}}^{(\rho^*,l)}_{\boldsymbol{k}}(\boldsymbol{G}) =  V^{Z}_{(\rho^*,l)}(\boldsymbol{G})\,.
\end{aligned}
\end{equation}
} 

Indeed, specializing to the atomic subspace of a $\mathcal{C}_2$-symmetric sublattice $\rho^*_j$ and orbital $l$, Eq.\;(\ref{eq_shift_criterium}) simplifies to (thanks to the block-diagonal structure of Eq.\;(\ref{eq_blockdiagonal_U2}))
\begin{equation}
\widetilde{\overline{V}}^{(\rho^*_j,l)}_{\boldsymbol{k}}(\boldsymbol{G}) =  
V^{Z}_{(\rho^*_j,l)}(\boldsymbol{G}) = e^{\text{i} \boldsymbol{G}\cdot \boldsymbol{r}_{\rho^*_j} }\, \mathbb{1}_{n_l\times n_l}  = \pm  \mathbb{1}_{n_l\times n_l}\,,
\end{equation}
where we used Eq.\;(\ref{eq_c2sym_shift}). In the case of a non-$\mathcal{C}_2$-symmetric sublattice pair, $\{\boldsymbol{r}_{\rho_{2j-1}},\boldsymbol{r}_{\rho_{2j}}\}$, we find, again from Eq.\;(\ref{eq_blockdiagonal_U2}),
\begin{equation}
\widetilde{\overline{V}}^{(\{\rho_{2j-1};\rho_{2j}\},l)}_{\boldsymbol{k}}(\boldsymbol{G}) =  \mathbb{1}_{2n_l\times 2n_l} \,.
\end{equation}

The real Wilson loop in atomic flag limit in this case will similarly be given by the shift matrix:
\begin{equation}
    \widetilde{\overline{\mathcal{W}}}_{ [\boldsymbol{k}_0+\boldsymbol{G}\leftarrow \boldsymbol{k}_0]} = \widetilde{\overline{V}}_{\boldsymbol{k}}(\boldsymbol{G})
    \,.
\end{equation}
Given that the reciprocal lattice vector $\boldsymbol{G}$ was chosen as the shortest vector of the commensurate direction $\boldsymbol{G}/\vert \boldsymbol{G}\vert$, let us define the shortest direct lattice vector $\boldsymbol{R}_{\boldsymbol{G}} = m_1\bm{a}_1 + m_2\bm{a}_2$ ($m_1,m_2\in \mathbb{Z}$) that is dual to $\boldsymbol{G}$, i.e. $e^{\text{i}\, \boldsymbol{G}\cdot \boldsymbol{R}_{\boldsymbol{G}} }=1$. From the above equation it is clear that for the subspace of all the degrees of freedom located at $\mathcal{C}_2$-centers, the Zak phase of an individual atomic orbital site $\gamma[\boldsymbol{k}_0+\boldsymbol{G}\leftarrow \boldsymbol{k}_0] = 0$ physically corresponds to a centered orbital, i.e. located at the site $[\bm{r}=0]$, and $\gamma=\pi$ physically corresponds to an  off-centered orbital, i.e. located at the site $[\bm{r}=\frac{\boldsymbol{R}_{\boldsymbol{G}}}{2}]$. We thus conclude that the non-$\mathcal{C}_2$-symmetric sublattice sites, and the degrees of freedom located at a centered $\mathcal{C}_2$-center, can be represented by a periodic Bloch basis (in the flag atomic limit) with a real-valued Bloch Hamiltonian matrix, while the degrees of freedom located at off-centered $\mathcal{C}_2$-centers cannot, at the same time, be represented by a periodic Bloch basis and realize a real-valued Bloch Hamiltonian matrix. We note the special case when all the degrees of freedom are located on off-centered $\mathcal{C}_2$-centers, such that $V^Z(\boldsymbol{G}) = -\mathbb{1}$, for which the Bloch Hamiltonian matrix can be brought to a periodic and real-valued form by changing the BL origin (see Section~\ref{sec:BLorigin_gauge} below).

\subsection{Periodictization Procedure and 1D Topological Atomic Obstructions of $\mathcal{C}_2\mathcal{T}$-symmetric Euler Systems}\label{sec:BLorigin_gauge}

We have considered above the various choices of basis that govern the periodicity of the real Bloch Hamiltonian matrix of 2D $\mathcal{C}_2\mathcal{T}$-symmetric systems under the assumption that a fixed BL origin was chosen. We now address the effect of changing the BL origin on the shift matrix, establish which BL origin choice is natural for Euler systems, and motivate how, while the choice of BL origin determines the apparent Zak phases, the physical interpretation and connection to the 1D topology of the system is unchanged.

To see how the overall phase of $V(\bm{G})$, for a reciprocal lattice vector $\bm{G} = m \bm{b_1} + n \bm{b_2}$ ($m,n \in \mathbb{Z}$), amounts to a shift of the Bravais lattice(BL) origin, we note that changing the global phase modifies the entries of the phase matrix~\eqref{eq:BZBC&Vmat_defn} by:
\begin{align}
    V(\bm{G}&)\to e^{i\Delta_{\bm{G}}} \ V(\bm{G})\nonumber \\
    \implies& e^{i\bm{G}\cdot \bm{r}_{\alpha}} \to e^{i\bm{G}\cdot \bm{r}_{\alpha}+\Delta_{\bm{G}}}\equiv e^{i\bm{G}\cdot \overbrace{(\bm{r}_{\alpha}+\bm{\delta})}^{\equiv \Tilde{\bm{r}}_{\alpha}}},
\end{align}
where we have defined an overall shift of the atomic sites $\bm{\delta}$ using $\Delta_{\bm{G}} = \bm{G} \cdot \bm{\delta}$. Thus, it is clear that a shift of the BL origin by $-\bm{\delta}$ applies a phase $e^{i\Delta_{\bm{G}}}$ to $V(\bm{G})$.

In the main text we have considered atomic lattices for which we could define a BL where all atomic orbitals lie at $\mathcal{C}_2$ centers, which we refer to as \textit{$\mathcal{C}_2$-centered lattice}. We note that the $\mathcal{C}_2$ centers are the following four equivalence class of sites for any 2D BL: $[\bm{r} = 0]$,$[\bm{r} = \frac{\bm{a}_1}{2}]$,$[\bm{r} = \frac{\bm{a}_2}{2}]$, and $[\bm{r} = \frac{\bm{a}_1+\bm{a}_2}{2}]$. As discussed in Section.~\ref{subsubsec:eumeration-of_possible_lattices}, it follows that four possible BL origin choices exist for $\mathcal{C}_2$-centered lattices. We pick the BL origin choice that gives an even number of off-centred orbitals along each of the primitive lattice vector directions in the direct lattice, i.e. the one for which $\det(V(\bm{b_1})) = \det(V(\bm{b_2})) = +1$, as it provides a natural choice to examine the effect of node braiding on the Euler topology since the corresponding atomic flag limit Dirac strings occur for a pair of bands simultaneously. More precisely, one can always order the Wannier states basis~\eqref{eq:wannierbasis} such that the $-1$ entries in the shift matrices are successive so that atomic flag limit Dirac strings occur for a pair of adjacent bands, much like how they manifest between nodes in Euler phases~\cite{Bouhon2020_NatPhys_ZrTe}. However, such ordering is not necessary as long as there is an even number of Bloch eigenvectors that are anti-periodic in the atomic flag limit (have a minus entry in their shift matrix component ) since a Dirac string between two non-adjacent bands $i$ and $j$ is equivalent to the composition of all the two-band sub-space Dirac strings enclosed in between the bands.  

We note that the aforementioned BL origin choice always exists for $\mathcal{C}_2$ centered lattices with an odd number of atomic orbitals (i.e. odd number of bands), but for those with an even number of atomic orbitals (i.e. even number of bands) it may or may not exist depending on how the atomic orbitals are positioned relative to each other [one can readily verify this by considering how atomic orbitals are split across the four distinct $\mathcal{C}_2$ centers while accounting also for the four possible BL origin choices in each case.]

We emphasize that the choice of BL origin is purely a gauge choice that does not affect the Euler topology or the edge states of the system and indeed one can check that while picking the gauge choice where either or both of the phase matrices have a negative determinant, which results in single band Dirac strings (that cannot be undone for all bands by node braiding), changes the apparent Zak phases, it leads to the same physical topological phases, i.e. Euler class values and edge states, upon non-Abelian braiding of nodes. This can be readily confirmed upon examining an example model and upon noting that the edge states are correctly accounted for by comparing the Zak phases of a given Euler phase with the that of the atomic limit of the model in the same BL origin gauge choice.

\section{Meronic Hamiltonian Model}   \label{AppSec:MeronH}
We provide explicit equations for the meronic Hamiltonian for the Kagome lattice presented in Fig.~\ref{Fig:MeronModel}a) in the main text and which is obtained from~\cite{Meron_paper}. In particular we have that the atomic site positions $\bm{r}_{\alpha}$ are given by~\cite{Meron_paper}:
\begin{equation}\label{Eq:r_a-positions}
    \bm{r}_A = \frac{\bm{a}_1}{2},\ \bm{r}_B = \frac{\bm{a}_2}{2},\ \bm{r}_C = -\frac{\bm{a}_1+\bm{a}_2 }{2},
\end{equation}
where $\bm{a}_1= \frac{3}{2}\hat{\mathbf{x}} + \frac{\sqrt{3}}{2}\hat{\mathbf{y}}$, $\bm{a}_2=- \frac{3}{2}\hat{\mathbf{x}} + \frac{\sqrt{3}}{2}\hat{\mathbf{y}}$ , and the lattice constant is chosen to be unity. The neighbor bond vectors are given by~\cite{Meron_paper}:
\begin{align}
    \bm{\delta}_{AB} = -\bm{r}_C, \ \bm{\delta}_{AC} = -\bm{r}_B, \ \bm{\delta}_{BC} = -\bm{r}_A \nonumber \\
    \bm{\delta}^\prime_{AB}= \bm{r}_B - \bm{r}_A, \ \bm{\delta}^\prime_{AC}= \bm{r}_A - \bm{r}_C, \ \bm{\delta}^\prime_{BC}=\bm{r}_C - \bm{r}_B  \nonumber\\
    \bm{\delta}^{\prime \prime}_{AA}=2\bm{r}_A , \ \bm{\delta}^{\prime \prime}_{BB}= 2\bm{r}_B, \ \bm{\delta}^{\prime \prime}_{CC}= 2\bm{r}_C.
\end{align}
\section{Quench Dynamics of Euler Systems}\label{AppSec:Quench}
We provide further details regarding the Hopf linking signatures that arise upon quenching with BZ non-periodic Euler Hamiltonians. First to obtain the relations for the transformation of the linking patterns across the BZBCs for an initial state $\bm{\Psi}_0$ such as Eq.~\eqref{Eq:p_relations} in the main text, we note the dependence of the projected Bloch vector $\bm{p}(\bm{k},t)$ on the Bloch eigenstate $\bm{n}(\bm{k})$ through $H_{\text{flat}}$~\eqref{Eq:H_flat}, which in turn appears in the expression of the evolved state: $\bm{\Psi}(\bm{k},t) = \left[ \text{cos}(t)- i \ \text{sin}(t)H_{\text{flat}}(\bm{n}(\bm{k})) \right]\cdot \bm{\Psi}_0$ through the Rodrigues form~\cite{Even_Quench}. The said relations could be then obtained through applying the BZBC $\bm{n}(\bm{k} + \bm{G}) = V(\bm{G})\ \bm{n}(\bm{k})$ followed by algebraic manipulations to obtain an expression for $\bm{p}(\bm{k}+\bm{G},t)$.

Analogous to the relations we obtain in the main text for $\bm{\Psi}_0= \hat{\bm{x}}$, we obtain the following relation for $\bm{\Psi}_0= \hat{\bm{y}}$:
\begin{align}\label{Eq:p_relationsY}
    \bm{p}(\bm{k}+\bm{G}, t) = 
    \begin{cases}
        \bm{p}(\bm{k}, t) & \text{if $V(\bm{G})=v_0$}, \\ \\
        -v_2 \cdot \Tilde{\bm{p}}(\bm{k}, t) & \text{if $V(\bm{G})=v_1$}, \\ \\
         -v_3 \cdot \Tilde{\bm{p}}(\bm{k}, t) & \text{if $V(\bm{G})=v_2$}, \\ \\
         v_1 \cdot \bm{p}(\bm{k}, t)  & \text{if $V(\bm{G})=v_3$},
    \end{cases}
\end{align}
and likewise for $\bm{\Psi}_0= \hat{\bm{z}}$;
\begin{align}\label{Eq:p_relationsZ}
    \bm{p}(\bm{k}+\bm{G}, t) = 
    \begin{cases}
        \bm{p}(\bm{k}, t) & \text{if $V(\bm{G})=v_0$}, \\ \\
        v_1 \cdot \bm{p}(\bm{k}, t) & \text{if $V(\bm{G})=v_1$}, \\ \\
         -v_2 \cdot \Tilde{\bm{p}}(\bm{k}, t) & \text{if $V(\bm{G})=v_2$}, \\ \\
         -v_3 \cdot \Tilde{\bm{p}}(\bm{k}, t) & \text{if $V(\bm{G})=v_3$},
    \end{cases}
\end{align}
upon making use of the corresponding $\boldsymbol{\mu}$ matrices presented in Appendix C of Ref.~\cite{Even_Quench}.

\end{document}